\documentclass[aps,prd,twocolumn,footinbib]{revtex4-1}
\usepackage{epsfig}
\usepackage[colorlinks,linkcolor=blue,anchorcolor=blue,citecolor=blue,urlcolor=blue,breaklinks=true]{hyperref}
\usepackage{amsmath}
\usepackage{booktabs} 
\usepackage{multirow}
\usepackage{array}
\usepackage{graphics}
\usepackage{slashed}
\usepackage{color}
\usepackage{amsmath}
\usepackage{bm}
\usepackage{setspace}
\usepackage{comment}
\emergencystretch=\maxdimen
\begin{document}

\author{He-Xia Zhang}
\email{zhanghexia@mails.ccnu.edu.cn}
\address{ Key Laboratory of Quark \& Lepton Physics (MOE) and Institute of 	Particle Physics, Central China Normal University, Wuhan 430079, China}
\author{Yu-Xin Xiao} %
\address{ Key Laboratory of Quark \& Lepton Physics (MOE) and Institute of 	Particle Physics, Central China Normal University, Wuhan 430079, China}
\author{Jin-Wen Kang} 
\address{ Key Laboratory of Quark \& Lepton Physics (MOE) and Institute of 	Particle Physics, Central China Normal University, Wuhan 430079, China}

\author{Ben-Wei Zhang}
\email{bwzhang@mail.ccnu.edu.cn}
\affiliation{Key Laboratory of Quark \& Lepton Physics (MOE) and Institute of  Particle Physics, Central China Normal University, Wuhan 430079, China} 
\affiliation{Guangdong Provincial Key Laboratory of Nuclear Science, Institute of Quantum Matter, South China Normal University, Guangzhou 510006, China.}

\title{Phenomenological study of the anisotropic quark matter  in  the 2-flavor Nambu-Jona-Lasinio model }

\begin{abstract}
	  With the two flavor Nambu-Jona-Lasinio (NJL) model we carry out a phenomenological study on the chiral phase structure, mesonic properties and   transport  properties in a momentum-space anisotropic quark matter. To calculate transport coefficients we have utilized  the  kinetic theory in the relaxation time approximation, where  the momentum anisotropy  is embedded  in  the estimation of both distribution function and the relaxation time. It is shown that  an increase of the anisotropy parameter $\xi$ may results in a catalysis of chiral symmetry breaking. The critical endpoint  (CEP) is shifted to smaller temperatures and larger quark chemical potentials  as $\xi $ increases, the impact  of momentum anisotropy on temperature of  CEP is almost the same as that on the quark chemical potential of CEP.
The meson masses and the associated decay widths  also exhibit a significant $\xi$ dependence.
	It is observed that   the  temperature behavior of  scaled shear viscosity $\eta/T^3$ and scaled electrical conductivity $\sigma_{el}/T$
	  exhibit a similar dip structure,    with the minima of both $\eta/T^3$  and  $\sigma_{el}/T$   shifting  toward higher temperatures with increasing $\xi$. Furthermore, we demonstrate that  the  Seebeck coefficient $S$ decreases when temperature goes up and its sign  is positive, indicating  the dominant carriers for converting  the temperature gradient to the electric field are up-quarks.
	    The Seebeck coefficient $S$ is significantly enhanced with a large  $\xi$ for the temperature below the critical temperature.
\bigskip

\end{abstract}

\maketitle

\section{introduction}

The properties of strongly interacting matter described by the quantum chromodynamics (QCD) in extreme conditions of temperature $T$ and density have aroused a plethora of experimental and theoretical studies in the last thirty years. The   experiment    studies performed 
at the Relativistic Heavy Ion Collider (RHIC)  in BNL and the Large Hadron Collider (LHC) in CERN   have  revealed that a  new deconfined state of  matter $-$  the quark-gluon plasma (QGP), can  be created  at high temperature  and/or baryon chemical potential $\mu_B$. And  lattice QCD  calculation, which is a powerful gauge invariant approach to investigate the non-perturbative properties, also  has confirmed   that  the phase transition  is a smooth and continuous crossover for vanishing  chemical potential~\cite{Bazavov:2014pvz,Aoki:2006we,Borsanyi:2013bia,Borsanyi:2010cj,Bernard:2004je,Aoki:2006br}.  
Due to the so-called fermion sign problem~\cite{Splittorff:2007ck}, lattice QCD  simulation  is  limited  to low finite density~\cite{Barducci:1989wi,Asakawa:1989bq,Bazavov:2018mes},  eventhough several calculation techniques  such as  the Taylor expansion~\cite{talor1,talor2},  analytic continuations from imaginary to real chemical potential~\cite{imagnary1,imagnary2}, multi-parameter reweighting method~\cite{Fodor:2001au} have been proposed to tackle this problem and  improve the validity  at high chemical potential. More detailed review of lattice calculation can be found in Refs.~\cite{lattice1,lattice2}. 
Therefore,  for arbitrary $\mu_B$ one has to rely on effective models to study the QCD phase transition.
Currently, there are  various QCD inspired  effective models such as  the Nambu-Jona-Lasinio (NJL) model~\cite{3flavor-NJL,NJL1,NJL2,Buballa:2003qv}, the 
Polyakov-loop enhanced NJL (PNJL) model~\cite{Ratti:2006wg,Mukherjee:2006hq,Costa:2010zw,Fukushima:2008wg},  the Quark-Meson (QM) model~\cite{Schaefer:2006ds,Schaefer:2008hk,Tripolt:2013jra}, the Polyakov QM (PQM) model~\cite{PQM,Skokov:2010wb,Schaefer:2011ex}, which not only  can successfully describe the spontaneous symmetry breaking and  restoration of QCD but also  have been applied to explore QCD phase structure  and internal properties of meson at arbitrary $T$ and $\mu_B$. 	
 And  these
 models  calculations  have predicted  (see e.g.~\cite{Schaefer:2008hk,Zhuang:1994dw})
   that at high chemical potential, the phase transition is a first-order phase transition, and with decreasing $\mu_B$,  the  first-order phase transition  has to end at a critical end point (CEP)   and change into a crossover. At this CEP  the  phase transition is of second order. However, due to various approximation adopted in the model  calculations, there is not an agreement on the existence and location of CEP on the phase diagram.
 Furthermore, the effects of rotation~\cite{Zhang:2020hha,Jiang:2016wvv}, the magnetic field effects~\cite{Gatto:2010pt,Kashiwa:2011js,DElia:2018xwo,Andersen:2013swa,Ferreira:2015jrm,Bali:2011qj}, finite-volume  effects~\cite{Wan:2020vaj,Zhao:2019ruc,Palhares:2009tf,Liu:2020elq,Xu:2020loz,Magdy:2019frj,XiaYongHui:2019gci,Tripolt:2013zfa,Bhattacharyya:2012rp,Deb:2020qmx}, non-extensive effects~\cite{Zhao:2020wks,Shen:2017etj,Rozynek:2009zh,Ishihara:2019ran}, external electric fields~\cite{Tavares:2019mvq,Ruggieri:2016xww,Cao:2015dya,Ruggieri:2016lrn}, and the effects of chiral chemical potential~\cite{Shi:2020uyb,Braguta:2016aov,Yu:2015hym,Lu:2016uwy}  also have been considered in the effective models   to  provide a better insight in  the  phase transition of the  realistic QCD plasma.

Apart from the importance of QCD phase structure information, the transport coefficients, characterizing  the non-equilibrium dynamical evolution of QCD matter, also  have captured large attention.
The shear viscosity $\eta$, 
 which  quantifies the rate of momentum transfer in  the fluid with inhomogeneous flow velocity, has been successfully used in the viscous relativistic hydrodynamic description of the QGP bulk dynamics. 
The small shear viscosity to entropy density ratio $\eta/s$ can  be exctrated from  the   elliptic flow data~\cite{v2}.  
In the literature, there are  various frameworks  for estimating $\eta$ of strongly interacting matter, e.g., the kinetic theory within the relaxation time approximation (RTA),  the  QCD effective models~\cite{Ghosh:2014vja,Marty:2013ita,shear-NJL,shear-PNJL,Zhuang:1995uf,Rehberg:1996vd},
 the  quasiparticle model (QPM)~\cite{shear-quasi1,shear-quasi2},  lattice QCD simulation~\cite{shear-Lattice5}, ect.
The electrical conductivity $\sigma_{el}$, as the response of a medium  to an applied electric field,  also has  attracted more attention  in high energy physics due to the presence of strong electromagnetic field created in the early stage of non-central  heavy-ion collisions  (HICs). The presence of $\sigma_{el}$   not only  can affect the duration and strength  of magnetic fields~\cite{McLerran:2013hla,Gursoy:2014aka}, but also    is directly proportional to the emissivity and production of  soft photon~\cite{Gupta:2003zh,Yin:2013kya}.
The thermal behavior of $\sigma_{el}$  has been estimated  using different approaches, such as the microscopic transport models~\cite{Hammelmann:2018ath,Steinert:2013fza,Greif:2014oia,electrical-phsd}, lattice gauge theory simulation~\cite{electrical-Lattice1,electrical-Lattice2,electrical-Lattice4,Amato:2013naa},  hadron resonance gas  model~\cite{shear and electrical,electrial2,electrical-EVHRG}, quasiparticle models~\cite{Mykhaylova:2020pfk,Bluhm:2009ef}, the effective models~\cite{Marty:2013ita,Soloveva:2020hpr}, the string percolation model~\cite{electrical-string1,electrical-string2}, the holographic method~\cite{Jain:2010ip,Finazzo:2013efa}  and so on.  Recently, the studies of  electrical conductivity  in the  QGP  at   magnetic fields  have also  been performed~\cite{Thakur:2019bnf,Kurian:2017yxj,Rath:2020idp,electrical1,Astrakhantsev:2019zkr}.
Another less concerned but interesting coefficient is  Seebeck coefficient (or called thermopower).  When a spatial  gradient of temperature   exists in a conducting medium, a corresponding electric field   can  arise and vice versa, which is the Seebeck effect.  When the  electric current induced by electric field  can compensate with  the current of temperature gradient, the thermal diffusion ends.   Accordingly,  the efficiency of converting temperature gradient to electric field in the  open circuit  condition is  quantified by   Seebeck coefficient $S$.  
 In past years, the Seebeck effect  has been  extensively investigated in condensed matter physics. Very recently, some exploration  has been  extended to the QCD matter. For example,  Seebeck coefficient  at magnetic fields and at  zero magnetic field  has been studied   in  both the  hadronic matter~\cite{Das:2020beh,Bhatt:2018ncr} and the QGP~\cite{Zhang:2020efz,Dey:2020sbm,Kurian:2021zyb}. In Ref.~\cite{Abhishek:2020wjm},  Seebeck coefficient also has been estimated based  the NJL model, where the  spatial gradient of quark chemical potential  is  considered apart from the presence of  temperature gradient.

In the beginning of HICs,  the pressure gradient of created fireball along the beam direction (denoted as longitudinal $z$ direction) is greatly lower than along transverse direction. After the rapid expansion of medium along the beam direction, the system becomes much colder in the beam direction than the transverse direction~\cite{Baier}, which causes  the QGP possess  the local momentum anisotropy, and this anisotropy can survive in the entire evolution of medium~\cite{Strickland:2014pga}.
In addition, the presence of  strong magnetic field also can induce a local   anisotropy in  momentum-space.

Inspired by the presence of momentum-space anisotropy in HICs, 
the primary objective of present work is to   study phenomenologically 
the effect of  momentum anisotropy  induced by the  rapid longitudinal expansion of medium on the chiral phase structure, mesonic properties, and transport coefficients in the SU(2) NJL model.
To incorporate momentum anisotropy into numerical calculations, we follow the parametrization  method proposed by Romatschke and Strickland~\cite{Romatschke:2003ms}, where the   isotropic momentum-space distribution functions of particle are deformed by rescaling  one preferred   direction in momentum space  and  introducing a directional dependent parameter $\xi$. 	This method has been extensively employed 
	to phenomenologically   study the impacts of  momentum anisotropy  on various observables, such as  photon  production~\cite{Kasmaei:2019ofu,Schenke:2006yp,Bhattacharya:2015ada}, the parton self-energy~\cite{Kasmaei:2016apv,Kasmaei:2018yrr,Ghosh:2020sng}, heavy-quark potential~\cite{Nopoush:2017zbu,Dumitru:2007hy}, and various transport coefficients~\cite{Thakur:2017hfc,Rath:2019vvi,Srivastava:2015via,Baier:2008js}.  And the relativistic anisotropic hydrodynamics (aHydro) models can  give a higher accurate description of non-equilibrium dynamics compared to other hydrodynamical models~\cite{Alqahtani:2017mhy}. As done in most previous studies, the present focus  is on the weakly anisotropic medium (very close to the equilibrium) for which  $|\xi|\ll1$ and  the distribution function can be expanded up to linear order in $\xi$. We can find that even at the small values of anisotropy, the effective quark mass and meson masses change significantly  compared to  the equilibrium result.  
Unlike most momentum anisotropy studies of  transport coefficients in the QGP, where the effect of momentum anisotropy is  not considered into the  different particle interaction channels, in present work, we incorporate the  momentum anisotropy  to the estimation of  the relaxation time to better  study the impact of $\xi$ on transport properties of quark matter near the  phase transition temperature region.

 The paper is organized as follows: Sec.~\ref{sec:NJLmodel} gives a brief  review of the basic formalism of the 2-flavor NJL model.
 In Sec.~\ref{sec:Masst} and Sec.~\ref{sec:Mass}, we present the brief derivation of the expressions associated  with the  constituent quark mass  and meson mass spectrum  in an isotropic and  anisotropic medium, respectively.
  Sec.~\ref{sec:coefficients}
  includes a detailed procedure for  obtaining  the formulae of momentum-anisotropy-dependent  transport coefficients. In Sec.~\ref{sec:tau}, we present the estimation of the relaxation time for (anti-)quarks.
     The   numerical results for various  observables  are phenomenologically analyzed in  Sec.~\ref{sec:result}. In Sec.~\ref{sec:sum}, the present work is summarized with an outlook. The  formulae of the matrix elements squared  for different  quark-(anti-)quark elastic scattering processes  are given in the  Appendix.
\section{Theoretical frame}\label{sec:NJLmodel}
In this work, we start from  the standard
two-flavor NJL model, which is a purely fermionic theory due the absence of  all gluonic degrees of freedom. Accordingly, the lagrangian is given as \cite{NJL1}
\begin{eqnarray}
\mathcal{L}=\bar{\psi}(i\slashed{\partial}-\hat{m}_0)\psi+G[(\bar{\psi}\psi)^2+(\bar{\psi}i\gamma_{5}{\bf{\hat{\tau}}}\psi)^2],
\end{eqnarray}
where $\psi (\bar{\psi})$ stands for the  quark (antiquark) field with two-flavors ($u,d$) and three colors $N_{c}=3$. $\hat{m}_0$ denotes  the diagonal matrix of the  current quark mass of  $up$ and $down$-quarks, $\hat{m}_0=diag(m_{u}^0,m_{d}^0)$ and we take $m_0=m_{u}^0=m_{d}^0$ to ensure isospin symmetry of the NJL lagrangian. $G$ is the effective coupling strength of  four-point fermion interaction   in the scalar and pseudoscalar channels.
 $\hat{\tau}$ is the vector of Pauli matrix in the  isospin space.

	In the NJL model, 
	under  the mean field (or Hartree) approximation~\cite{NJL1,NJL2,3flavor-NJL}  the  quark   self-energy is momentum-independent and can be identified as the constituent quark mass $m_q$, which acts  as   order parameter for charaterizing chiral phase transition.
	 For an off-equilibrium system,  the evolution of space-time dependence of the constituent quark mass in the closed-time-path formalism can be obtained by solving  the  gap equation \cite{Zhang:1992rf}
\begin{eqnarray}\label{eq:Gap1}
m_q
=m_{0}-2Gi\mathrm{Tr}iS^{<}(x,x),
\end{eqnarray}
 where $S^{<}(x,y)=i\langle\bar{\psi}(y)\psi(x)\rangle$ with $x=(t,\mathbf{x})$  is real time Green function in coordinate space~\cite{Botermans:1990qi,Rehberg:1998nd}, $\langle\dots\rangle$  denotes the average over the ensemble under consideration, and the trace runs over spin, color and flavor degrees of freedom. 
Transforming Eq.~(\ref{eq:Gap1}) to phase space with the help of the Winger transformation, and introduce the quasiparticle approximation (see Ref.~\cite{Rehberg:1997it} for details), the gap equation  
can be  further written as~\cite{Zhang:1992rf,Rehberg:1997it,Klevansky:1997wm,Rehberg:1998nd} 
\begin{equation}\label{eq:Gap2}
m_q=m_0+4N_fN_c\int\frac{d^3\mathbf{p}}{(2\pi)^3}\frac{m_q}{E_{\mathbf{p}}}\bigg(1-f_{q}(x,\mathbf{p})-\bar{f}_{\bar{q}}(x,\mathbf{p})\bigg),
\end{equation}
where $E_{\mathbf{p}}=\sqrt{\mathbf{p}^2+m_q^2(x)}$ is the quasi-quark energy. Since the NJL model is a non-renormalizable model due to the point-like four fermion interaction in the lagrangian, an ultraviolet cutoff $\Lambda$ is used to regularize the divergent integral. 
In the  non-equilibrium case, the space-time  evolution of one-particle distribution function $f(x,\mathbf{p})$ in  Eq.~(\ref{eq:Gap2}) is described by  Boltzmann-Vlasov transport equation of NJL model in Hartree level~\cite{Klevansky:1998rs,Klevansky:1997wm,Wang:2020wwm}. 
 By  solving the Vlasov equation together with the gap equation cocurrently,  the constituent quark mass affecting the space-time dependence of $f(x,\mathbf{p})$ can be determined self-consistently.

 To better understand the meson dynamics in HICs, it's useful to study the structure of meson propagation in the medium.
 In the framework of NJL model,  mesons  are  quark-antiquark bound states or collective modes. The meson propagator  can be  constructed by calculating the quark-antiquark effective scattering  amplitude within the random phase approximation (RPA)~\cite{NJL2,Hatsuda:1994pi,Zhang:1992rf,Rehberg:1998nd}.
Following   Refs.~\cite{Zhang:1992rf,Rehberg:1998nd}, the  explict form for the pion ($\pi$)  and the  sigma meson ($\sigma$) propagators in the  RPA  are as follows
 \begin{eqnarray}\label{eq:Dt}
 D_{M}(x,k_0,\mathbf{k})=\frac{2iG}{1-2G\Pi^{}_{M}(x,k_0,\mathbf{k})}.
 \end{eqnarray}
All information of meson is contained in the irreducible one-loop pseudoscalar or scalar  polarization function $\Pi^{}_{M}$ with 
 the  subscript  $M$ corresponding to pseudoscalar ($\pi$) or scalar ($\sigma$) mesons. 
 The space-time dependence of  polarization function in an off-equilibrium system    is given as~\cite{Zhang:1992rf,Rehberg:1998nd}
 \begin{eqnarray}\label{eq:PI-NO}
 \Pi_{M}&=&N_cN_f\int\frac{d^3\mathbf{p}}{4\pi^3}\frac{1}{E_{\mathbf{p}}}[1-f_q(x,\mathbf{p})-f_{\bar{q}}(x,\mathbf{p})]\nonumber\\
 &&\times\biggl\{1-\frac{(k_0^2-\mathbf{k}^2-\nu_{M}^2)}{2E_{\mathbf{p}-\mathbf{k}}}\bigg[\frac{E_{\mathbf{p}}+E_{\mathbf{p}-\mathbf{k}}}{k_0^2-(E_{\mathbf{p}}+E_{\mathbf{p}-\mathbf{k}})^2}\nonumber\\
 &&-\frac{E_{\mathbf{p}}-E_{\mathbf{p}-\mathbf{k}}}{k_0^2
 	-(E_{\mathbf{p}}-E_{\mathbf{p}-\mathbf{k}})^2}\bigg]\biggr\}.
\end{eqnarray}
Here, $\nu_M$ denotes the real value of the bound meson energy, and   $\nu_{M}=0 ~(2m_q)$ for  $\pi$ $(\sigma)$ meson.  
\section{Constituent quark and meson  in   an isotropic  quark matter }\label{sec:Masst}
In an expanding system (e.g., the dynamical process of heavy-ion collisions), the space-time dependence in distribution function is hidden in the space-time dependence of  temperature and chemical potential. However, for a uniform temperature and chemical potential, i.e., for a  system in  global equilibrium, the distibution function is well defined and  independent of space-time.
  Therefore, in the  equilibrium (isotropic) state, to  investigate  chiral phase transition  and mesonic properties within  NJL model,   one can  employ imaginary-time formalism.  
  Actually, the  results in Ref.~\cite{Zhang:1992rf} have indicated that   the real-time calculation of the closed time-path Green's function  reproduces exactly the finite temperature result of the NJL model obtained from  Matsubara's temperature Green's function in the themodynamical equilibrium limit. 
  In the following, we will briefly give some procedure for the derivation of polarization function  in imaginary time formalism.
In the  equilibrium system,  $m_q$, which  is temperature- and quark chemical potential- dependent, can be directly calculated from the self-consistent  gap equation in momentum space~\cite{NJL2,3flavor-NJL,NJL1}:
\begin{equation}\label{eq:Gap}
m_q=m_0+4GN_fN_c\int\frac{d^3\mathbf{p}}{(2\pi)^3}\frac{m_q}{E_{\mathbf{p}}}\bigg(1-f_{q}^0(\mathbf{p})-\bar{f}^0_{\bar{q}}(\mathbf{p})\bigg).
\end{equation}
We can see that by taking thermal  equilibrium   distribution functions (Fermi-Dirac distributions)  Eq.~(\ref{eq:Gap2})  is the same form as Eq.~(\ref{eq:Gap}).
The   equilibrium   distribution function of (anti-)quark $f_{q(\bar{q})}^0$ can be given as  
\begin{eqnarray}
f^{0}_{q(\bar{q})}(\mathbf{p})=[\exp[(E_\mathbf{p}- \mu_{q(\bar{q})})\beta]+1]^{-1},
\end{eqnarray}
where $\beta=1/T$ is the  inverse  temperature of system.  And an uniform quark chemical potential $\mu\equiv\mu_{u,d}\equiv-\mu_{\bar{u},\bar{d}}$ is assumed. It's note that  the ultraviolet divergence is not presented in the integrand  containing Fermi-Dirac distribution functions, so the momentum integral don't need to be regularized for finite temperatures.
In the equilibrium, the meson propagator is given as~\cite{Zhuang:1995uf}
 \begin{eqnarray}\label{eq:D}
D_{M}(k_0,\mathbf{k})=\frac{2iG}{1-2G\Pi^{}_{M}(k_0,\mathbf{k})},
\end{eqnarray}
where $\Pi_{M}^{}$ at an arbitary  temperature  and  quark chemical potential is given by~\cite{Rehberg:1995kh,Rehberg:1997xe}
\begin{eqnarray}\label{PI}
\Pi^{}_{M}(k_0,\mathbf{k})&=&-\frac{N_c N_f}{8\pi^2}[((m_q\mp m_q)^2-k_{0}^2+\mathbf{k}^2)\nonumber\\
&&\times B_{0}(k_{0},\mathbf{k},\mu,T,m_q)+2A(\mu,T,m_q)].
\end{eqnarray}
The minus (plus) sign  refers to   pseudoscalar (scalar)  mesons.   
The function $A$, which  relates to  the  one-fermion-line integral,   in   the imaginary time  formalism for finite temperatures and quark chemical potentials is given  as~\cite{Rehberg:1995kh}
\begin{equation}\label{eq:A}
A=\frac{16\pi^2}{\beta}\sum_{n}^{}\exp(i\omega_{n}y)\int\frac{d^3\mathbf{p}}{(2\pi)^3}\frac{1}{(i\omega_{n}+\mu)^2-E_{\mathbf{p}}^2},
\end{equation}
where $\omega_{n}=(2n+1)\pi/\beta$ are the fermionic Matsubara frequencies and the sum of $n$ runs over all positive and negative integer values. It is to be understood that the limit $y\rightarrow 0$ is to be taken after the Matsubara summation.
The function $B_0$  in Eq.~(\ref{PI}) relates to  the  two-fermion-line integral. At  finite $\mu$ and $T$, $B_0$ is defined as~\cite{Rehberg:1995kh}
\begin{eqnarray}\label{eq:B0}
&B_{0}&
(i\nu_l,\mathbf{k},\mu,T,m_q)\nonumber\\
&=&\frac{16\pi^2}{\beta}\sum_{n}^{}\exp(i\omega_{n}y)\int_{|\vec{p}|<\Lambda}\frac{d^3\mathbf{p}}{(2\pi)^3}\frac{1}{((i\omega_{n}+\mu)^2-E_{}^2)}\nonumber\\
&&\times\frac{1}{((i\omega_{n}-i\nu_{l}+\mu)^2-E_{}'^2)},
\end{eqnarray}
where we have abbreviated $E'= E_{\mathbf{p}-\mathbf{k}}=\sqrt{(\mathbf{p}-\mathbf{k})^2+m_q^2}$ and $E=E_{\mathbf{p}}$ for convenience, and   after the Matsubara summation on $n$ is carried out, the  complex  frequencies $i\nu_{l}$ are   analytically continued to their values on  the real plane, i.e., $i \nu_{l}\rightarrow k_{0}+i\epsilon$ ($\epsilon>0$) with   $k_0$ being the zero component of associated  four  momentum.
For the  full calculations of functions $A$ and $B_0$ at  arbitrary values of temperature and chemical potential can be  found in  Ref.~\cite{Rehberg:1995nr}.
After evaluating the Matsubara summation   by contour integration in the usual fashion~\cite{Fetter}, 
Eq.~(\ref{eq:A}) is given as 
\begin{eqnarray}\label{eq:AT}
A=8\pi^2\int\frac{d^3\mathbf{p}}{(2\pi)^3 E_\mathbf{p}}\big(f_{q}^{0}+f^{0}_{\bar{q}}-1\big),
\end{eqnarray}
where the Fermi-Dirac distribution function is introduced.
Similar to the treatment of  the function A,  after Matsubara summation over $n$ and the Matsubara frequencies $i\nu_{l}$  are  analytically continued  to real values, the  Eq.~(\ref{eq:B0}) can be rewritten as  the  following form:
\begin{eqnarray}\label{eq: B0 general1}
&B_{0}&(k_0,\mathbf{k},\mu,T,m_q)\nonumber\\
&=&16\pi^2\int_{}\frac{d^3\mathbf{p}}{(2\pi)^3}\frac{f_q^{0}(p)+f_{\bar{q}}^{0}(p)-1}{2E2E'}\nonumber\\
&&\times\bigg[\frac{1}{E+k_0-E'+i\epsilon}-\frac{1}{E+k_0+E'+i\epsilon}\nonumber\\
&&+\frac{1}{E-k_0-E'-i\epsilon}-\frac{1}{E-k_0+E'-i\epsilon}\bigg].
\end{eqnarray}
Inserting Eqs.~(\ref{eq:AT}-\ref{eq: B0 general1}) to Eq.~(\ref{PI}),  we finally can  obtain the  expression of $\Pi_{M}$  in the equilibrium state, which is  formally the same  as Eq.~(\ref{eq:PI-NO}),  except that  the distribution functions are  ideal Fermi-Dirac distribution function  rather than space-time dependent distribution functions.

\section{Constituent quark and meson  in   a weakly anisotropic medium  }\label{sec:Mass}
As aforemetioned in the introduction, the consideration of momentum anisotropy induced by  rapid expansion of the hot QCD medium  for  existing phenomenological  applications is mostly achieved by parameterizing  associated  isotropic   distribution functions. 
To proceed the numerical  calculation, a specific  form of      anisotropic (non-equilibrium)  momentum distribution function is required. 
In this work, we utilize the  Romatschke-Strickland (RS) {\color{blue}form}~\cite{Romatschke:2003ms} in which  the system  exhibits  a spheroidal momentum anisotropy, and the anisotropic distribution is obtained from an arbitrary  isotropic distribution function   by removing particle with a momentum component along  the direction of anisotropy, $\mathbf{n}$. As done in the  literature~\cite{Dumitru:2009fy}, we restrict ourselves here to a plasma close to equilibrium and so the isotropic functions are   thermal equilibrium distributions, viz, Fermi-Dirac distributions.
Accordingly, the explicit form of  anisotropic  momentum distribution function in the local rest frame can be given   as
\begin{equation}\label{eq:covariantfan}
f^{an}(\mathbf{p})=\frac{1}{\exp[(\sqrt{\mathbf{p}^2+\xi (\mathbf{p}\cdot\mathbf{n})^2+m_q^2}- \mu'_{q(\bar{q})})\beta']+1},
\end{equation}
It is  worth noting that for  the anisotropic matter, the  $T'$ and $\mu' $ appearing in Eq.~(\ref{eq:covariantfan})  lose the usual meaning of  $T$ and $\mu$ in the equilibrium system and become
	dimensionful scales related to the mean particle momentum~\cite{Romatschke:2006bb}, which is due to the fact that in the presence of anisotropy the system is away from equilibrium. If we assume the system to be very close to the equilibrium (in the  small anisotropy limit) then the parameters $T'$  and $\mu'$ still could be taken to be $T$ and $\mu$ respectively,  as done in Ref.~\cite{Chandra:2010xg}.
The anisotropy parameter $\xi$ reflects 
the degree of anisotropy and is defined as  $\xi=\langle p_{T}^2\rangle/(2\langle p_{L}^2\rangle)-1$,  $p_{T}=|\mathbf{p}-(\mathbf{p}\cdot\mathbf{n})\cdot\mathbf{n}|$ and $p_{L}=\mathbf{p}\cdot\mathbf{n}$ are the  momentum components of particles perpendicular and parallel to $\mathbf{n}$, respectively. 
	 Since the precise time evolution of $\xi $ is still an open question,  therefore, the 
	 anisotropy parameter $\xi$ in local  anisotropic system is restricted to be constant and independent  of time.
$-1<\xi<0$ corresponds to a contraction of  momentum  distribution along the direction of anisotropy and  $\xi>0$ corresponds to a stretching of momentum distribution along the direction of anisotropy.
In Eq.~(\ref{eq:covariantfan}), the  three-velocity of partons and anisotropy unit vector    are choosen as 
\begin{eqnarray}
\mathbf{n}&=&(\sin \chi,0,\cos\chi),\\
\mathbf{p}&=&p(\sin\theta\cos\phi,\sin\theta\sin\phi,\cos\theta),
\end{eqnarray}
where $\chi$ is the angle  between $\mathbf{p}$ and $\mathbf{n}$, and $p\equiv |\mathbf{p}|$ throughout the computations.
Within this choice,  the spheroidally anisotropic term, $\xi(\mathbf{n}\cdot\mathbf{p})^2$, in Eq.~(\ref{eq:covariantfan}) can be   written as
$\xi(\mathbf{n}\cdot\mathbf{p})^2=\xi p^2(\sin\chi\cos\phi\sin\theta+\cos\chi\cos\theta)^2=\xi c(\theta,\phi,\chi)$.
We further  assume $\mathbf{n}$  points  along the beam ($z$) axis, i.e., $\mathbf{n}=(0,0,1)$.
 It is essential to note that  we shall restrict ourselves here to a plasma close to equilibrium state   and  has small anisotropy around equilibrium state.
 Therefore, in  the weak anisotropy limit ($|\xi|\ll 1$), one can expand   Eq.~(\ref{eq:covariantfan}) around the isotropic limit and retain only the leading order  in  $\xi$. Accordingly,  the  anisotropic  momentum distribution function in local rest frame can be further written as~\cite{Srivastava:2015via}
\begin{eqnarray}\label{eq:fan}
f^{an}(\mathbf{p})= f^{0}-\frac{\xi(\mathbf{n}\cdot\mathbf{p})^2}{2E_{}T} f^{0}(1-f^{0}),
\end{eqnarray}
where the second term  is the 
    anisotropic correction to equilibrium distribution, which  is also related to the  leading-order viscous correction to equilibrium distribution in  viscous hydrodynamics.
For a fluid expanding one-dimensionally  along the direction $\mathbf{n}$  in the Navier-Stokes limit,  the explicit relation  is given as~\cite{Asakawa:2006jn,Dumitru:2009ni}
\begin{eqnarray}
\xi=\frac{10}{T\tau}\frac{\eta}{s},
\end{eqnarray} 
which indicates that non-zero shear viscosity (finite momentum relaxation rate) in an expanding system also  can explicitly  lead to the presence of  momentum-space anisotropy.
  At the RHIC energy with the critical temperature $T_{c}\approx160$~MeV, $\tau\approx6 $ fm/c and $\eta/s=1/4\pi$, we can obtain $\xi\approx 0.3$.
In principle, for the  non-equilibrium dynamics of the chiral phase transition, a self-consistent numerical study must be performed by solving the Boltzmann-Vlasov  transport equation  together with gap   equation in terms of  space-time dependent quark distribution as mentioned in Sec.~\ref{sec:NJLmodel}. However,  due to that the non-equilibrium distribution function in the local rest frame of  weakly anisotropic systems 
	 has a specific form  and the temperature and chemical potential appearing in  Eq.~(\ref{eq:fan}) are still considered as  free parameters, i.e., the space-time evolution  is  not addressed, as done in the literature~\cite{Chandra:2010xg}.
Therefore,	in  the non-equilibrium states possessing small momentum space anisotropy, just by solving  gap equation with  anisotropic momentum distribution i.e., Eq.~(\ref{eq:fan}), we can phenomenologically investigate the impact of    momentum anisotropy on the temperature and quark potential dependence of constituent quark mass. 
Accordingly,   the gap equation, i.e.,  Eq.~(\ref{eq:Gap1}) or Eq.~(\ref{eq:Gap}) can be modified as
\begin{eqnarray} 0
&=&\bigg[2N_cN_fG\int_{0}^{\infty}\frac{p^2dp}{\pi^2}\frac{m_q}{E}\bigg(-f^0_{q}-f^0_{\bar{q}}+1+\frac{p^2\xi F_p}{6E_{}T}\bigg)\bigg]\nonumber\\
&&+m_0-m_q.
\end{eqnarray}
Here we have abbreviated $F_{p}=f^0_q(1-f^0_q)+f^0_{\bar{q}}(1-f^0_{\bar{q}})$ for convenience.
The momentum anisotropy  also is embedded in  the study of mesonic properties 	by substituting the  anisotropic momentum distribution in the $A$ function part of Eq.~(\ref{eq:PI-NO}), one obtains 
\begin{eqnarray}
A
=4\int\frac{p^2 dp}{E}\bigg[f^0_{q}+f^0_{\bar{q}}-1-\frac{\xi p^2 F_{p}}{6ET}\bigg].
\end{eqnarray}
In the $\xi\to 0$ limit, above equation  reduces to Eq.~(\ref{eq:AT}).
Similar to the treatment of  function A, the  weak momentum anisotropy effects  can enter the $B_0$ function (as given in  Eq.~(\ref{eq:PI-NO}) or Eq.~(\ref{eq: B0 general1}))  by  taking the   anisotropic momentum distribution.
Without the loss of generality, we  choose the coordinate system in such a way that $\mathbf{k}$ is parallel to  $z$-axis, i.e.,
\begin{eqnarray}
\mathbf{k}&=&(0,0,k), \quad|\mathbf{k}|\equiv k.
\end{eqnarray}
We first discuss a simple case, i.e., $\mathbf{k}=\mathbf{0}$,  $k_0\neq0$, the computation of  function $B_{0}$  in a weakly anisotropic medium  is trivial, $viz$,
\begin{eqnarray}\label{eq: B0 for k=0}
B_{0}&=&\pi^2\int\frac{d^3\mathbf{p}}{(2\pi)^34E^2}(f^{an}_{q}(\mathbf{p})+f^{an}_{\bar{q}}(\mathbf{p})-1)\nonumber\\
&&\times(-\frac{1}{k_0+2 E-i\epsilon}-\frac{1}{-k_0+2 E+i\epsilon}).
\end{eqnarray}
In the integrand of the  above equation, there are two  poles  at $E=E_0=\pm k_{0}/2$ if $m\leq E_{0}$.
Applying   the Cauchy formula
\begin{equation}
\mathrm{lim}_{\epsilon\to 0}\frac{1}{x-i\epsilon}=\mathcal{P}\frac{1}{x}+i\pi\delta(x),
\end{equation}
where $\mathcal{P}$ denote the Cauchy principal value, finally  function $B_0$ can be rewritten as 
\begin{eqnarray}
B_{0}^{}&=&
8\mathcal{P}\int\frac{p^2dp}{E_{}(k_0^2-4E^2)}[f^{0}_{q}(\mathbf{p})+f^{0}_{\bar{q}}(\mathbf{p})-1-\frac{\xi p^2}{6ET}F_{p}]\nonumber\\
&&-i\pi\frac{2}{k_0}\bigg[\bigg(f^{0}_q(z_0)+f^{0}_{\bar{q}}(z_0)-1-\frac{\xi z_0^2}{3k_{0}T}F_{z_0}\bigg)\nonumber\\
&&\times\sqrt{(\frac{k_0}{2})^2-m_q^2}\Theta(\frac{k_0^2}{4}-m_q^2)\bigg].
\end{eqnarray}
Here, $z_{0}=\sqrt{(\frac{k_0}{2})^2-m_q^2}$,  and $\Theta$ is the step function to ensure the imaginary part appears only for $k_{0}/2>m_q$.  
Next, in the case of $k>0,~k_0\neq 0$,  the expression of function $B_0$  is slightly complicated and  can be  written as 
\begin{eqnarray}\label{eq:case3}
B_{0}&=&
\frac{1}{k}\int_{0}^{\infty}\frac{pdp}{E}(f_{q}^{an}(\mathbf{p})+f_{\bar{q}}^{an}(\mathbf{p})-1))\nonumber\\
&&\times\int_{-1}^{1}dx\big(\frac{1}{x+(k_0^2+2k_0 E-k^2)/2pk-i\epsilon}\nonumber\\
&&+\frac{1}{x+(k_0^2-2k_0 E-k^2)/2pk+i\epsilon}\big)\nonumber\\
&&={B}_{0}^{an}+B_{0}^{iso},
\end{eqnarray}
with the abbreviation  $x=\cos\theta$.
The isotropic part ${B}_{0}^{iso}$ reads as
\begin{eqnarray}
B_{0}^{iso}
&=&\frac{1}{k}\int_{0}^{\infty}\frac{pdp}{E}\bigg(f^0_q+f^0_{\bar{q}}-1\bigg)\nonumber\\
&&\times\bigg[\log\bigg|\frac{(k_0^2-k^2+2pk)^2-(2k_0 E)^2}{(k_0^2+k^2-2pk)^2-(2k_0 E)^2}\bigg|\nonumber\\
&&+i\pi(\Theta(2pk-|k_0^2+2k_0 E-k^2|)\nonumber\\
&&-\Theta(2pk-|k_0^2-2k_0 E-k^2|))\bigg].
\end{eqnarray}
And the anisotropic part ${B}_{0}^{an}$ reads  as
\begin{widetext}
	\begin{eqnarray}
	{B}_{0}^{aniso}&=&-\frac{1}{k}\int_{0}^{\infty}\frac{pdp}{E}
	\frac{\xi p^2 F_p}{2ET}\bigg[\bigg(-\frac{(k_0^2-k^2+2k_0 E)}{pk}+\bigg(\frac{k_0^2-k^2+2k_0 E}{2pk}\bigg)^2\log\bigg|\frac{k_0^2+2k_0 E-k^2+2pk}{k_0^2+2k_0 E-k^2-2pk}\bigg|\nonumber\\
	&&-\frac{(k_0^2-k^2-2k_0 E)}{pk}+\bigg(\frac{k_0^2-k^2-2k_0 E}{2pk}\bigg)^2\log\bigg|\frac{k_0^2-2k_0 E-k^2+2pk}{k_0^2-2k_0 E-k^2-2pk}\bigg|\bigg)\nonumber\\
	&&+i\pi\bigg(\frac{k_0^2+2k_0 E-k^2}{2pk}\Theta(2pk-|k_0^2+2k_0 E-k^2|)-\frac{k_0^2-2k_0 E-k^2}{2pk}\Theta(2pk-|k_0^2-2k_0 E-k^2|)\bigg)\bigg].
	\end{eqnarray}
\end{widetext}
When  $k_0=0$,  the  imaginary part of Eq.~(\ref{eq:case3}) vanishes. Therefore,   the real and imaginary parts of meson polarization functions for different cases  in  a weakly anisotropic medium are given as  
\begin{widetext}
	\begin{eqnarray}\label{eq:Re(vec0)}
	\mathrm{Re} \Pi_M(k_0,0)=N_fN_c\mathcal{P}\int_{0}^{\infty}\frac{dpp^2}{\pi^2 E}\bigg[1-f^0_q(\mathbf{p})-f^0_{\bar{q}}(\mathbf{p})+\frac{\xi p^2}{6ET}F_p\bigg]\frac{E^2-\nu_{M}^2/4}{E_{}^2-(k_0/2)^2},
	\end{eqnarray}
	\begin{eqnarray}\label{eq:Im(vec0)}
	\mathrm{Im}\Pi_M(k_0,0)=\frac{N_cN_f}{8
		\pi k_0}\sqrt{k_0^2-4m_q^2}(k_0^2-\nu_{M}^2)\bigg[1-f^0_q(z_0)-f^0_{\bar{q}}(z_0)+\frac{z_0^2\xi}{3k_0T}F_{z_{0}}\bigg]\Theta(k_0^2-4m_q^2),
	\end{eqnarray}
	\begin{eqnarray}\label{eq:Re(vec)}
	Re\Pi_M(0,k)&=&\frac{N_cN_f}{\pi^2}\int_{0}^{\infty}\frac{dp p^2}{E}(1+\frac{k^2+\nu_{M}^2}{4pk}\ln\bigg|\frac{k-2p}{k+2p}\bigg|)\bigg(1-f^0_q(\mathbf{p})-f^0_{\bar{q}}(\mathbf{p})\bigg)\nonumber\\
	&&+\frac{N_cN_f}{\pi^2}\int_{0}^{\infty}\frac{\xi p^2dp}{E^2T}F_p\bigg[\frac{p^2}{6}+\frac{k^2+\nu_{M}^2}{4}\bigg(1+\frac{k^2}{4pk}\ln\bigg|\frac{k-2p}{k+2p}\bigg|\bigg)\bigg].
	\end{eqnarray}
\end{widetext}
When the effect of momentum anisotropy is turned off  ($\xi=0$),   Eqs.~(\ref{eq:Re(vec0)})-(\ref{eq:Re(vec)}) are reduced to the results of Ref.~\cite{Zhuang:1995uf} in thermal equilibrium.
Once the propagators of mesons are given, their masses then can be  determined by the pole in Eq.~(\ref{eq:D}) at zero three- momentum~\cite{Deb:2016myz}, i.e.,
\begin{eqnarray}\label{eq:solution}
1-2G \mathrm{Re}\Pi_{M}(m_{\pi,\sigma},0)=0.
\end{eqnarray}
The solution  is  real value  for $m_{\pi,\sigma}<2m_{q}$,  a meson  is stable. However, for $m_{\pi,\sigma}>2m_{q}$, a meson  dissociates to its constituents and becomes a resonant state. Accordingly, the polarization function is a complex function and $\Pi_{M}$ has an imaginary part that is related  to the decay width of the resonance as $\Gamma_{M}=\mathrm{Im}\Pi_{M}(m_{\pi,\sigma},0)/m_{\pi,\sigma}$~\cite{Deb:2016myz}.
\section{Transport coefficients in an anisotropic quark matter}\label{sec:coefficients}
In this section, we start to  study the effects due to  a local  anisotropy of the plasma in momentum space on the  transport coefficients (shear viscosity,  electrical conductivity and    Seebeck coefficient) in  quark matter. The calculation is performed in the kinetic theory that  is widely used to  describe the   evolution of the  non-equilibrium  many-body system in the dilute limit.
Assuming that the system  has a slight deviation from the equilibrium, the relaxation time approximation (RTA) is reasonably employed.  The momentum anisotropy  is encoded in the phase-space distribution function which evolves according to the relativistic Boltzmann equation. 
We give the procedures of deriving the  $\xi$-dependent transport coefficients below.
\subsection{  Shear viscosity}
   The propagation of  single-quasiparticle   whose mass is temperature- and chemical potential- dependent in the anisotropic medium is described by the relativistic Boltzmann-Vlasov equation~\cite{Plumari:2012ep,Klevansky:1998rs}
\begin{eqnarray}\label{boltzmann}
\bigg[p^{\mu}\partial_{\mu}+\frac{1}{2}\partial^\mu m_a^2\partial_\mu^{(p)}\bigg] f_a(x,\mathbf{p})=C[f_a(x,\mathbf{p})],
\end{eqnarray}
where $\frac{1}{2}\partial^\mu m_a^2$ acts as the force term attributed to the residual mean field interaction. 
 The right-hand side of Eq.~(\ref{boltzmann})  is the collision term. 
Considering the system has  a small  departure from the equilibrium due to external perturbation,
 the collision term within the RTA can be given as,
\begin{equation}\label{eq:Cf}
C[f]\simeq - 
\frac{p^\mu u_{\mu}[f_{a}(x,\mathbf{p})-f_{a}^{0}(x,\mathbf{p})]}{\tau_{a}}=-\frac{p^\mu u_{\mu}\delta f_a}{\tau_{a}},
\end{equation}
 in which  $\tau_{a}$ denotes the relaxation time for particle species $a$, and can   quantify how fast the system reaches the equilibrium again. 
  The late-time equilibrium  distribution function of particle species $a$  is given as
 \begin{eqnarray}
 f^0_a(x,\mathbf{p})=[\exp((\mu_{\nu}(x)p^\nu- \mu_a(x))\beta(x))+1]^{-1},
 \end{eqnarray}
 where $p^{\nu}\equiv(E_{a},\mathbf{p})$  is particle   four-momentum,  $u^{\nu}=\gamma_\nu(1,\mathbf{u})$ is fluid four-velocity with $\gamma_\nu =(1-\mathbf{u}^2)^{1/2}$.
$\delta f_a$ in Eq.~(\ref{eq:Cf}) is the deviation of distribution function from  the local  equilibrium due to external disturbance, which  up to first-order in gradient expansion can reads as
\begin{eqnarray}\label{eq:deltafp}
\delta f_a=-\frac{\tau_{a}}{p^\mu u_{\mu}}\bigg[p^\mu \partial_{\mu} f_a^{an}+m_a\frac{d m_a}{d T}(\partial^\mu T)\partial_{\mu}^{(p)} f_a^{an}\bigg],
\end{eqnarray}
where the four-derivative can be decomposed into $\partial_{\mu}\equiv \partial/\partial^\mu\equiv u_{\mu}D+\nabla_{\mu}$,  $D\equiv u^\mu\partial_{\mu}$ and  $\nabla_\nu\equiv\Delta^{\mu\nu}\partial_{\nu}$  respectively denote the  time derivative  and  spatial gradient  operator in the local rest frame.
 $g^{\mu\nu}=diag(1,-1,-1,-1)$ is the metric tensor,
$\Delta^{\mu\nu}=g^{\mu\nu}-u^\mu u^\nu$ is the projection operator orthogonal to $u^\mu$.
In the presence of weak momentum anisotropy, the associated covariant version of  anisotropic function  for $a$-th particle  $f^{an}_a$  can be written as~\cite{Florkowski:2012lba}
 \begin{eqnarray}
f^{an}_a(x,\mathbf{p})&=&\frac{1}{\exp[(\sqrt{(p_\nu u^\nu)^2+\xi (p_\nu V^\nu)^2}- \mu_{a})\beta]+1},\\
&\approx&f_a^0-\frac{\xi(p^\nu V_{\nu})^2}{2Tp^\nu u_{\nu} }f_a^0(1-f_a^0)\label{eq:cfan1},
\end{eqnarray}
where  $V^{\nu}=(0,\mathbf{n})$ is defined as  the  anisotropy direction.
 Employing Eq.~(\ref{eq:cfan1}) in in Eq.~(\ref{eq:deltafp}), $\delta f_a$ can decompose into two part
\begin{eqnarray}\label{eq:dec}
\delta f_a=\delta f_a^{iso}+\delta f_a^{an}.
\end{eqnarray}
The first term on the right hand side  is 
\begin{eqnarray}
\delta f_a^{iso}&=&\frac{\tau_a}{p^\mu u_{\mu}}f_a^0(1-f_a^0)\bigg[p^\mu p^\nu \beta (u_{\mu}Du_{\nu}+\nabla_{\mu}u_{\nu})\nonumber\\
&&+p^\mu(p^\mu u_{\mu}) (u_{\mu}D\beta+\nabla_{\mu}\beta)+ \frac{dm_a^2}{dT^2} DT\bigg].
\end{eqnarray}
And employing the motion of equation in ideal hydrodynamics and ideal thermodynamic relations, $\delta f_a^{iso}$ can be rewritten as 
\begin{eqnarray}
\delta f_a^{iso}&=&\frac{\tau_a}{p^\mu u_{\mu}} f^0_a(1-f^0_a)\bigg\{\frac{p^\mu p^\nu}{T}\sigma_{\mu\nu}\nonumber\\
&&+\bigg[((p^\mu u_{\mu})^2-T^2\frac{d m_a^2}{d T^2})c_s^2+\frac{1}{3}\Delta_{\mu\nu}p^\mu p^\nu\bigg]\frac{\theta}{T}
\bigg\},
\end{eqnarray}
with  $\theta=\partial_{\alpha} u^{\alpha}$  and $c_{s}^2$ being the expansion rate of the fluid and squared sound velocity in the medium, respectively.
 The   velocity  stress tensor  has  the usual definition:
$\sigma^{\mu\nu}=\frac{1}{2}\Delta^{\mu\alpha}\Delta^{\nu\beta}(\nabla_{\alpha} u_{\beta}+\nabla_{\beta}u_{\alpha}-\frac{1}{3}\Delta^{\mu\nu}\theta)$.
After tedious calculations, one can  obtain  the second term in Eq.~(\ref{eq:dec}):
\begin{eqnarray}\label{eq:deltaan}
\delta f_a^{an}&=&\frac{(p^\nu V_{\nu})^2\xi\beta}{2p^\nu u_{\nu}}(2f_a^0-1)\delta f_a^{iso}\nonumber\\
&&-\frac{(p^\nu V_{\nu})^2\xi}{2(p^\nu u_{\nu})^2}\bigg[p^\mu p^\nu \beta (u_{\mu}Du_{\nu}+\nabla_{\mu}u_{\nu})\nonumber\\
&&-p^\mu(p^\mu u_{\mu}) (u_{\mu}D\beta+\nabla_{\mu}\beta)- \frac{ dm_a^2}{dT^2} DT\bigg]\nonumber\\
&&\times\frac{\tau_a}{p^\mu u_{\mu}}f_a^0(1-f_a^0)\nonumber\\
&=&\frac{(p^\nu V_{\nu})^2\xi\beta}{2p^\nu u_{\nu}}(2f_a^0-1)\delta f_a^{iso}\nonumber\\
&&-\frac{(p^\nu V_{\nu})^2\xi}{2(p^\nu u_{\nu})^2}\frac{\tau_a}{p^\mu u_{\mu}}f_a^0(1-f_a^0)\bigg\{\frac{p^\mu p^\nu}{T}\sigma_{\mu\nu}\nonumber\\
&&-\bigg[((p^\mu u_{\mu})^2-T^2\frac{dm_a^2}{d T^2})c_s^2+\frac{1}{3}\Delta_{\mu\nu}p^\mu p^\nu\bigg]\frac{\theta}{T}\bigg\}.\nonumber\\
\end{eqnarray}
Allowing the system to be slightly out of equilibrium, 
the energy-momentum tensor $T^{\mu\nu}$ can be expanded as: $T^{\mu\nu}=T_{0}^{\mu\nu}+T_{diss}^{\mu\nu}$, where $T_{0}^{\mu\nu}$  is the ideal   perfect fluid form and $T_{diss}^{\mu\nu}$ is the dissipative part of the energy-momentum  tensor. 
In hydrodynamical description of hot QCD matter,   the dissipative part of energy-momentum tensor up to first order in the gradient expansion   has the following  form~\cite{Hosoya:1983xm}
\begin{eqnarray}\label{eq:T1}
T^{\mu\nu}_{diss}=\pi^{\mu\nu}-\Pi\Delta^{\mu\nu},
\end{eqnarray}
where $\pi^{\mu\nu}$ and $\Pi$ being the shear stress tensor and  bulk   viscous pressure, respectively. In present work,  our focus is  the  shear viscosity component only.
In the kinetic theory, the  first-order shear stress tensor $ \pi^{\mu\nu}$    can be constructed  in terms of the  distribution functions
\begin{eqnarray}\label{eq:T2}
\pi^{\mu\nu}=\int\frac{d^3\mathbf{p}}{(2\pi)^3}\frac{1}{u\cdot p}\Delta^{\mu\nu}_{\phi\gamma} p^\phi p^\gamma\delta f.
\end{eqnarray}
Here, the double projection operator  is defined as  $\Delta^{\mu\nu}_{\phi\gamma}=\frac{1}{2}(\Delta^\mu_{\phi}\Delta^\nu_{\gamma}+\Delta^\mu_{\gamma}\Delta^\nu_{\phi})-\frac{1}{3}\Delta^{\mu\nu}\Delta^{\phi\gamma}$, which  can project any rank-2 Lorentz tensor onto its transverse (to $u^\mu$) and traceless	part.
Inserting Eqs.~(\ref{eq:dec})-(\ref{eq:deltaan})  to Eq.~(\ref{eq:T2}) and comparing with  the first-order  Navier-Stokes equation $\pi^{\mu\nu}=2\eta \sigma^{\mu\nu}$~\cite{Landau},   in the  rest frame of thermal system with $u^\mu\equiv(1,\mathbf{0})$ and $p^\nu u_{\nu}=E_a$, we finally get the expression of $\xi$-dependent shear viscosity of $a$-th particle,
\begin{eqnarray}\label{eq:shear}
\eta_{a}&=&-\frac{\xi d_{a}}{180T^2}\int \frac{dp}{\pi^2}\frac{\tau_{a}p^8}{E_{a}^3}f_{a}^0(1-f_{a}^0)(1-2f_{a}^0+\frac{T}{E_a}
),\nonumber\\
&&+\frac{d_a}{30T}\int \frac{dp}{\pi^2}\frac{\tau_{a}p^6}{E_{a}^2}f_{a}^0(1-f_{a}^0),
\end{eqnarray}
which is consistent with the result from Ref.~\cite{Thakur:2017hfc}.  
For the system consisting of multiple  particle species, total shear viscosity   is given as  $\eta=\sum_{a}\eta_{a}$.  In SU(2) light quark matter, $a=u,d,\bar{u},\bar{d}$ and  the spin-color degeneracy factor reads explicitly  $d_a=2N_c$.

\subsection{ Electrical conductivity and  Seebeck coefficient }
We  also investigate the effect of momentum anisotropy on the electrical conductivity and the thermoelectric coefficient. Under the RTA, the relativistic Boltzmann-Vlasov equation for  the  distribution function of single-quasiparticle in the presence of  external electromagnetic field is given by 
\begin{eqnarray}\label{boltzmann2}
\bigg[p^{\mu}\partial_{\mu}+\bigg(\frac{1}{2}\partial^\mu m_a^2+qF^{\mu\nu}p_{\nu}\bigg)\partial_\mu^{(p)}\bigg] f_a=-\frac{p^\mu u_{\mu} \delta f_a}{\tau_a},\nonumber\\
\end{eqnarray}
where $F^{\mu\nu}$ is the electromagnetic field strength tensor.
We   only  consider the presence   of an external  electric  field,  $F^{i0}=-F^{0i}=\mathbf{\cal{E}} 
	=(\mathcal{E},0,0)$. It is convenient to work in  the local rest frame of plasma, and under steady state assumption ($f_a$ does not depend on time explicitly, $\frac{\partial f_a}{\partial t}$=0),
	 Eq.~(\ref{boltzmann2}) can be given by~\cite{Abhishek:2020wjm}
\begin{eqnarray}\label{eq:boltzmann}
\mathbf{v}_{a}\cdot\mathbf{\nabla} f_{a}+(e_{a}\mathbf{\mathcal{E}}-\nabla_{} E_a)\cdot\frac{\partial 
	f_{a}}{\partial 
	\mathbf{p}}=-\frac{\delta f_a}{\tau_a},
\end{eqnarray}
where we use of  the chain rule $\frac{\partial p^0}{\partial \mathbf{p}}\frac{\partial}{\partial p^0}+\frac{\partial}{\partial\mathbf{p}}\rightarrow \frac{\partial }{\partial \mathbf{p}}$.
 $e_{a}$ is the electric charge of  $a$-th particle. $\mathbf{v}_{a}=\partial E_{a}/\partial \mathbf{p}$ is velocity of     particle species $a$. 
In order to solve Eq.~(\ref{eq:boltzmann}), we assume the  deviation of distribution function in an anisotropic medium
satisfies the following linear form:
\begin{eqnarray}\label{eq:fa}
\delta f_{a}^{}=-\tau_{a}(e_a\mathbf{\mathcal{E}}-\frac{\partial E_a}{\partial \mathbf{x}}
)\cdot\frac{\partial 
	f_{a}^{an}}{\partial 
	\mathbf{p}}-\tau_{a}\mathbf{v}_a\cdot\frac{\partial 
	f_{a}^{an}}{\partial \mathbf{x}},
\end{eqnarray}
The spatial gradient of the equilibrium isotropic  distribution  $\partial_{\mathbf{x}}f^0_{a}$ in the presence of medium-dependent quasiparticle mass can be expressed as  the following linear form:
\begin{equation}
\partial_{\mathbf{x}}f^0_{a}=-f^0_{a}(1-f^0_{a})\left(\partial_{\mathbf{x}}(\frac{E_{a}}{T})-\partial_{\mathbf{x}}(\frac{\mu_{a}}{T})\right),
\end{equation}
where  $\mu_a=t_a\mu$  denotes quark chemical potential of $a$-th particle  and $t_a=+1(-1)$ for the quark (antiquark). Considering $\mu$  is homogeneous in space and   the  temperature gradient   only  exists along $x$-axis,  and 
inserting Eq.~(\ref{eq:fa}) into Eq.~(\ref{eq:boltzmann}),
consequently,  the  perturbative term $\delta f^{}_a$ in an anisotropic medium can be written as 
\begin{eqnarray}\label{eq:deltaf}
\delta f^{}_a
&=&H_a\tau_{a}(e_a\mathcal{E}v_x)
-G_a\tau_{a}\partial_{x}(\frac{\mu_a}{ T}) v_x\nonumber\\
&&-\frac{\xi(\mathbf{p}\cdot\mathbf{n})^2}{E_aT}f_a^0(1-f_a^{0})\tau_{a}\partial_x E_a.
\end{eqnarray}
The expressions of  $H_a$ and $F_a$ in above equation can respectively  read as 
\begin{eqnarray}
H_a&=&\frac{1}{T} f_a^0(1-f_a^{0})(1+\xi c(\theta,\phi,\chi))\nonumber\\
&&-\frac{\xi p^2c(\theta,\phi,\chi)}{2E_{a}T^2}f_a^{0}(1-f_a^{0})(1-2f_a^0+\frac{T}{ E_{a}}),\\
G_a&=&\frac{1}{T^2}f_a^{0}(1-f_a^{0})-\frac{\xi p^2 c(\theta,\phi,\chi)}{2E_{a}T^3}(E_a-\mu_{a})\nonumber\\
&&\times f_a^{0}(1-f_a^{0})(1-2f_a^{0}-\frac{T}{ E_a-\mu_{a}}).
\end{eqnarray}
In linear response theory, the general formula of electric current density  $\mathbf{J}_a$ for particle species $a$ in response to external  electric field ($\mathcal{E}$) and temperature gradient ($\nabla_{x}T$) is given by ~\cite{Jaiswal:2015mxa}
\begin{eqnarray}
\mathbf{J}_a=\sigma_{el,a}(\mathbf{\mathcal{E}}-S_a\nabla_{x}T),
\end{eqnarray}
where $\sigma_{el,a}$ and $S_a$ are electrical conductivity and Seebeck coefficient of $a$-th particle, respectively.
$\mathbf{J}_a$  in term of distribution function within the kinetic theory can be written as
\begin{eqnarray}
\mathbf{J}_a=e_ad_a\int\frac{d^{3}\mathbf{p}}{(2\pi)^{3}}\mathbf{v}_a\delta 
f_a.
\end{eqnarray}
Finally,  the expressions of   $\sigma_{el,a}$ and $S_a$  in the weakly anisotropic medium  are respectively obtained as,
\begin{eqnarray}\label{eq:sigma-xx-xy}
\sigma_{el,a}&=&\frac{e^2_{a}d_{a}}{6T}\int \frac{dp}{\pi^2}\frac{\tau_{a}p^4}{E_{a}^2}f_{a}^0(1-f_{a}^0)(1+\frac{\xi}{3})\nonumber\\
&-&\frac{\xi e_{a}^2 d_a}{36T^2}\int\frac{dp}{\pi^2}\frac{\tau_{a}p^6}{E_{a}^3}f_{a}^0(1-f_{a}^0)(1-2f_{a}^0+\frac{T}{E_{a}}),\nonumber\\
\end{eqnarray}
and  
\begin{eqnarray}\label{eq:Seebeck}
S_{a}
=&&\frac{1}{\sigma_{el,a}}\bigg[\frac{e_{a}d_a}{6T^2}\int\frac{dp}{\pi^2}\frac{\tau_ap^4}{E_{a}^2}(E_{a}-\mu_{a})
f_a^{0}(1-f_a^{0})\nonumber\\
&&-\frac{e_{a}d_a\xi}{36T^3}\int\frac{dp}{\pi^2}\frac{\tau_ap^6}{E_{a}^3}(E_{a}-\mu_{a})f_a^{0}(1-f_a^{0})\nonumber\\
&&\times(1-2f_a^{0}-\frac{T}{(E_{a}-\mu_{a})})\bigg]=\frac{\alpha_a}{\sigma_{el,a}},
\end{eqnarray}
where  $\alpha_a$ is the  thermoelectric conductivity of   $a$-th particle. In the isotropic limit $\xi\to0$, our reduced expressions in Eqs.~(\ref{eq:sigma-xx-xy})-(\ref{eq:Seebeck}) are  identical to the formulae in Refs.~\cite{Deb:2016myz,gavin,Hosoya:1983xm}.
In condensed physics, semiconductor can exhibit either electron conduction (negative thermopower) or hole conduction (positive thermopower). Total thermopower in a material with  different carrier types is  given as the sum of respective contributions   weighted by  respective electrical conductivity values~\cite{Thermoelectric}. Inspired by this,  total Seebeck coefficient
 in a medium composed of light  quarks and antiquarks   can be given as 
\begin{eqnarray}
S=\frac{\sum_{a} S_{a}\sigma_{el,a}}{\sum_{a}\sigma_{el,a}}=\frac{\sum_{a}\alpha_a}{\sum_{a}\sigma_{el,a}}=\frac{\alpha}{\sigma_{el}},
\end{eqnarray}
where the fractional electric charges of  $up$ and $down$ (anti-)quarks are given explicitly by $e_u=-e_{\bar{u}}=2e/3$ and $e_d=-e_{\bar{d}}=-e/3$.  The electric charge reads $e=(4\pi\alpha_s)^{1/2 }$ with the fine structure constant $\alpha_s\simeq1/137$.

\section{computation of the  relaxation time }\label{sec:tau}
To quantify the transport coefficients, one needs to specify the relaxation time. 
Different to the treatment of  our previous work~\cite{Zhang:2020zrv}, where the relaxation time in  the calculation of transport coefficients is crudely taken as a constant, in present work the more  realistic scattering processes  through  the exchange  of meson are encoded into the estimation of the   relaxation time.
\begin{figure}
	\includegraphics[width=0.5\textwidth]{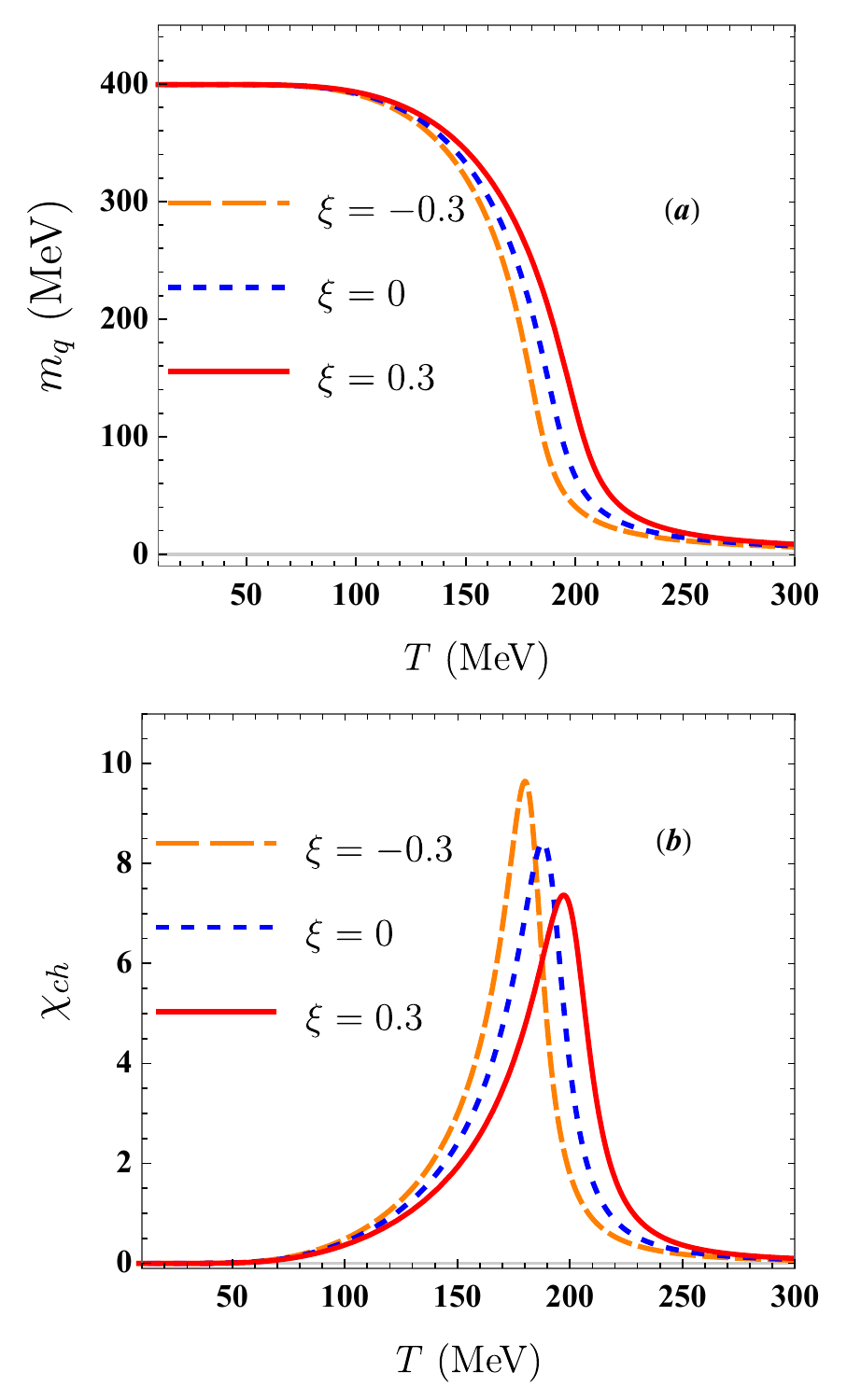}
	\caption{\label{Fig:Mass}  (plot a) The temperature dependences of  the constituent quark mass $m_q$ at  $\mu=0$~MeV for different fixed anisotropy parameters. (plot b)  The chiral susceptibility $\chi_{ch}$  at  $\mu=0$~MeV for different fixed anisotropy parameters.
		The broad dashed lines,  dashed lines and   solid lines   represent  to the results for $\xi=-0.3$, $\xi=0$ and 0.3, respectively. }
\end{figure}
The relaxation times  of   (anti)quarks are microscopically determined by the thermal-averaged  elastic scattering cross-section and the particle density. 
For light  quarks, the relaxation  time in the RTA   can be written  as~\cite{Zhuang:1995uf}
\begin{eqnarray}\label{eq:taulight}
\tau_{l}^{-1}(T,\mu)&=&n_{\bar{q}}[\bar{\sigma}_{u\bar{u}\rightarrow u\bar{u}}+\bar{\sigma}_{u\bar{u}\rightarrow d\bar{d}}+\bar{\sigma}_{u\bar{d}\rightarrow u\bar{d}}]\nonumber\\
&&n_{q}[\bar{\sigma}_{ud\rightarrow ud}+\bar{\sigma}_{uu\rightarrow uu}],
\end{eqnarray}
where the number density of (anti-)quarks in weakly aniosotropic medium is given as $ n_{q(\bar{q})}=d_{l}\int \frac{d^3p}{(2\pi)^3}f^{an}_{q(\bar{q})}$ with    $d_{l}=d_{\bar{l}}=2N_{c}$ denoting the degeneracy factor.
The momenta of the colliding  particles for the  elastic scattering  process $a(\mathbf{p}_1)+b(\mathbf{p}_2)\rightarrow c(\mathbf{p}_3)+d(\mathbf{p}_4)$ obey the relation $\mathbf{p}_1+\mathbf{p}_2=\mathbf{p}_3+\mathbf{p}_4=\mathbf{0}$, and  we use  the notation $|\mathbf{p}_1|=|\mathbf{p}_2|=p$ for convenience.
 In the center-of-mass (c.m.) frame, the mandelstam variables $s,~t,~u$~are defined as 
\begin{eqnarray}
s&=&4m_q^2+4p^2, t=-2p^2(1-\cos\theta_{p}) ,\nonumber\\
u&=&-2p^2(1+\cos\theta_{p}),
\end{eqnarray}
where $\theta_{p}$ is the scattering angle in the c.m. 
frame.
  The mandelstam variables hold the relation $ u+s+t=4m_q^2$.
 $\bar{\sigma}_{ab\rightarrow cd}$ denotes  energy-averaged  elastic  scattering cross-section  in the weakly anisotropic system, which  can be written  as 
\begin{eqnarray}\label{eq:cross section}
\bar{\sigma}_{ab\rightarrow cd}&=&\int_{s_{0}}^{\infty}ds\int_{t_{min}}^{t_{max}}dt\frac{d\bar{\sigma}_{ab\rightarrow cd}}{dt}\sin^2\theta_p
\int_{-1}^{1}dx_3\nonumber\\
&&\times(1-f_{c}^{an}(p_{cm},\mu,x_3))\int_{-1}^{1}	dx_4\nonumber\\
&&\times(1-f_{d}^{an}(p_{cm},\mu,x_4))\mathcal{L}(s,\mu,x_1,x_2),
\end{eqnarray}
with $(1-f_{c,d}^{an})$ denoting  the Pauli-blocking factor for the fermions due to the fact that some of the final states are already occupied by other identical  (anti-)quarks. 
 $\frac{d\bar{\sigma}}{dt}
=\frac{1}{16\pi s(s-4m_q^2)}|\bar{M}|^2$ is differential scattering  cross-section.
$|\bar{M}|^2$ denotes the matrix element squared of a specific scattering process.
 The formulae of  matrix elements squared for various scattering  processes are presented in the Appendix.
 The integration limits of $t$ are $t_{max}=0$ and 
$t_{min}=-4p_{cm}^2=-(s-4m_q^2)$ with $p_{cm}
=\sqrt{s-4m_q^2}/2$ denoting the momentum in the c.m. frame. The kinematic boundary of $s$ reads $s_{0}=4m_q^2$.  The scattering weighting factor $\sin^2\theta_p=\frac{-4t(s+t-4m_q^2)}{(s-4m_q^2)^2}$ is introduced to exclude the scattering processes with the small initial angle because  the large angle scattering is dominated in momentum transport process~\cite{Danielewicz:1984ww}.
 In the c.m. frame,  the leading-order anisotropic  distribution function  can be rewritten  as
\begin{eqnarray}
&f^{an}&(p_{cm},\mu,x)= f^{0}(p_{cm},\mu)\nonumber\\
&&-\frac{ p_{cm}^2\xi x^2}{2 E_{cm} T} f^{0}(p_{cm},\mu)(1-f^{0}(p_{cm},\mu)),
\end{eqnarray}
where  $E_{cm}=\frac{s-m_q^2+m_q^2}{2\sqrt{s}}=\sqrt{s}/2$.
In Eq.~(\ref{eq:cross section}), 
$\mathcal{L}$ denotes the probability of finding a quark-(anti)quark pair with the center of mass energy $\sqrt{s}$ in the  anisotropic medium, which is given as 
\begin{equation}
\mathcal{L}
=C\sqrt{s(s-4m_q^2)}f_a^{an}(p_{cm},\mu,x_1)f_b^{an}(p_{cm},\mu,x_2)v_{rel}(s),
\end{equation}
where  $v_{rel}(s)=\sqrt{\frac{s-4m_q^2}{s}}$ is the relative velocity between two incoming  particles in the c.m. frame; 
 $C$ is the normalization constant, and is determined  from the requirement that  
$\int_{s_{0}}^{\infty} ds\int_{-1}^{1} dx_1\int_{-1}^{1} dx_2 \mathcal{L}=1$ with abbreviations $x_1=\cos \theta_1$ and $x_2=\cos\theta_2$,  $\theta_i$  is  the angle between $\mathbf{p}_i$ and $\mathbf{n}$ and  $x_1\equiv-x_3$,~$x_2\equiv-x_4$.
Applying  the  above formula of  relaxation time to   Eqs.~(\ref{eq:shear}), (\ref{eq:sigma-xx-xy}), (\ref{eq:Seebeck}), we can calculate the transport coefficients  in the QCD medium and study their sensitivity to the momentum anisotropy.

\begin{figure}
	\includegraphics[width=0.45\textwidth]{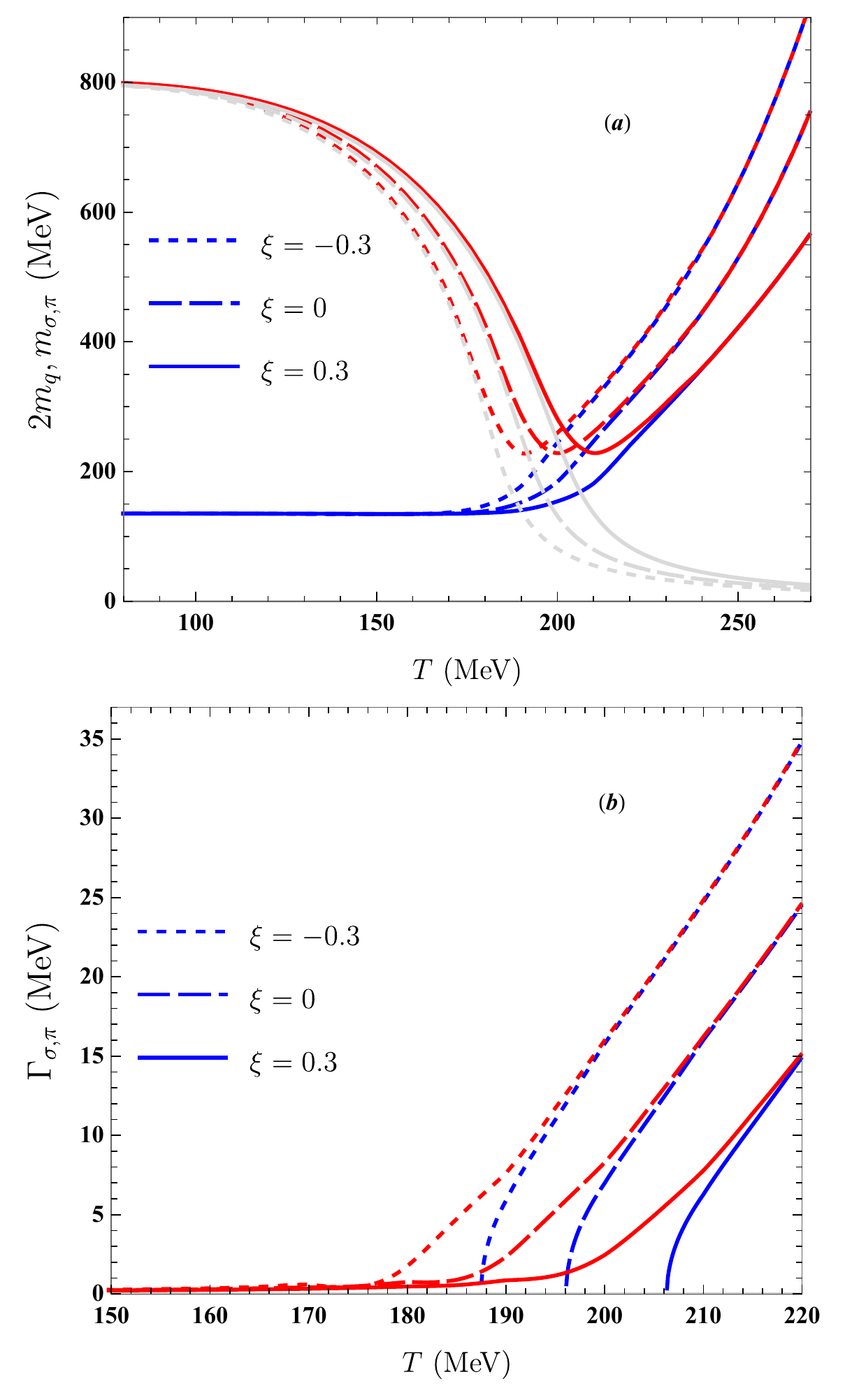}
	\caption{\label{Fig:Meson}   (plot a) The double mass of   constituent quarks $2m_q$ (gray lines), $\pi$ meson mass (blue lines) and $\sigma$ meson mass (red lines) as a function with temperature  at  $\mu=0$~GeV for different anisotropy parameter $\xi$.
		(plot b) The temperature dependences of both  $\pi$ (blue lines) and $\sigma$ (red lines) meson decay widths for different $\xi$.
The broad dashed lines,  dashed lines and   solid lines   represent  to the results for $\xi=-0.3$, $\xi=0$, and 0.3, respectively. The respective Mott temperatures are approximately 187~$\mathrm{MeV}$,~196~$\mathrm{MeV}$, 206~$\mathrm{MeV}$ for $\xi=-0.3,~0,~0.3$.	}
\end{figure}

\section{Results and discussion}\label{sec:result}
 Throughout this work, the following parameter set is used:  $m_{0}=m_{0,u}=m_{0,d}=5.6~\mathrm{MeV}$, $G\Lambda^2=2.44$ and $\Lambda=587.9~\mathrm{MeV}$. These values are taken from Ref.~\cite{Buballa:2003qv}, where these parameters are determined by fitting quantities  in the vacuum ($T=\mu=0$~MeV). At $T=0$, the chiral symmetry is spontaneously broken and one obtain the  current  pion mass $m_{0,\pi}=135$~MeV,  the pion decay constant $f_{\pi}=92.4$~MeV,  the  quark condensate $-\langle\psi\psi\rangle^{1/3}=241$~MeV. 
 
In the NJL model,  the constituent  quark mass  is a good indicator and an order parameter for  analyzing the  dynamical feature of  chiral symmetry.
 In the asymptotic expansion-driven momentum anisotropic system, the anisotropy  parameter $\xi$   is always  positive due to the rapid expansion along the  beam direction. Whereas, in the strong magnetic field-driven momentum anisotropic system, $\xi$  is   always negative due to the reduction of transverse momentum in Landau quantization. Since we restrict the analysis to only weakly anisotropic medium, the anisotropy parameters we work here  are artificially taken as $\xi=-0.3,~0.0,~ 0.3$, to phenomenologically investigate the effect of  $\xi$ on various quantities.
  In  Fig.~\ref{Fig:Mass} (a), we show the thermal behavior of light   constituent quark mass $m_{q}$  for  vanishing quark  chemical potential at different $\xi$. For low temperature, $m_{q}$ remains approximately constant ($m_q\approx$ 400 MeV), then  with increasing temperature  $m_{q}$ continuously drops to near zero. The transition to small mass occurs at higher temperature  for higher value of  $\xi$. These  phenomena imply that at zero chemical potential  the restoration of the chiral symmetry (the chiral symmetry  is not strictly restored)  because the current quark mass is nonzero) in   an   (an-)isotropic quark matter  takes place  as crossover phase transition, and an increase in $\xi$ can lead a catalysis of chiral symmetry breaking.
   \begin{figure*}\centering
  	\includegraphics[width=1.0\textwidth]{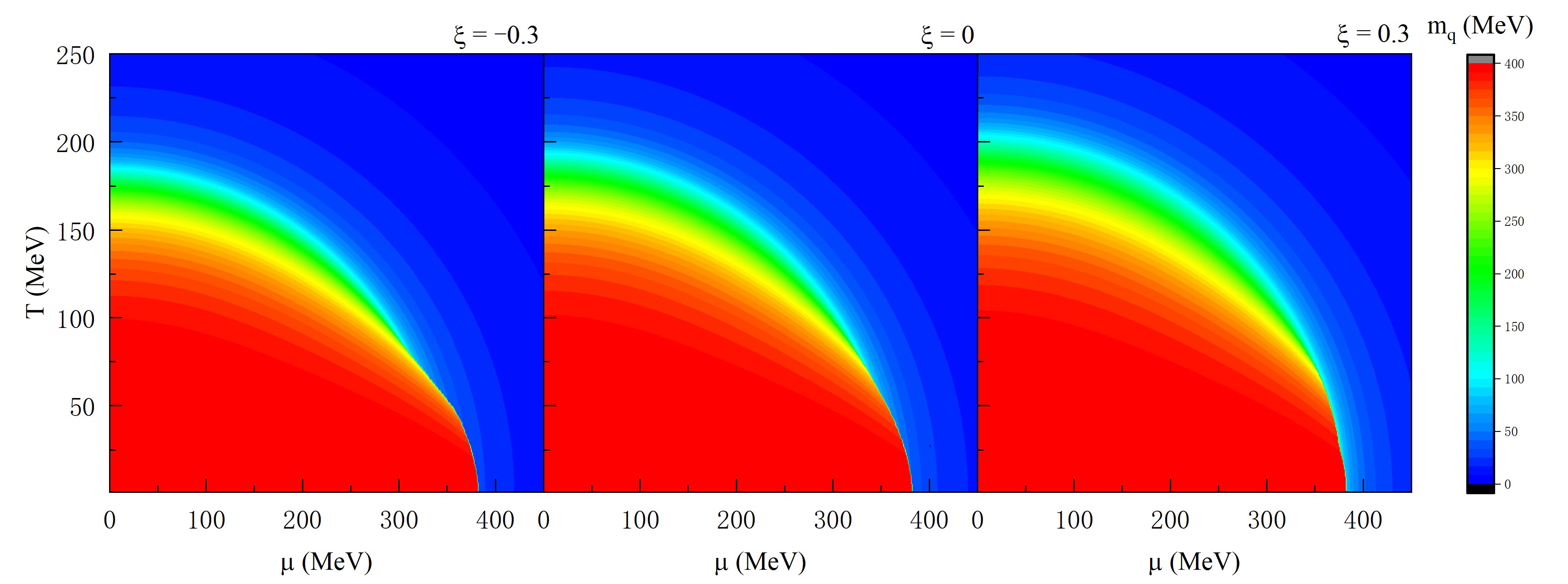}
  	\caption{\label{Fig:mq_uT} The 3-dimensional plot of constituent quark mass $m_q$ with respect to temperature  and quark chemical potential for different anisotropy parameters ($\xi=-0.3,~0,~0.3$). }
  \end{figure*}

 \begin{figure}\centering
 	\includegraphics[width=0.5\textwidth]{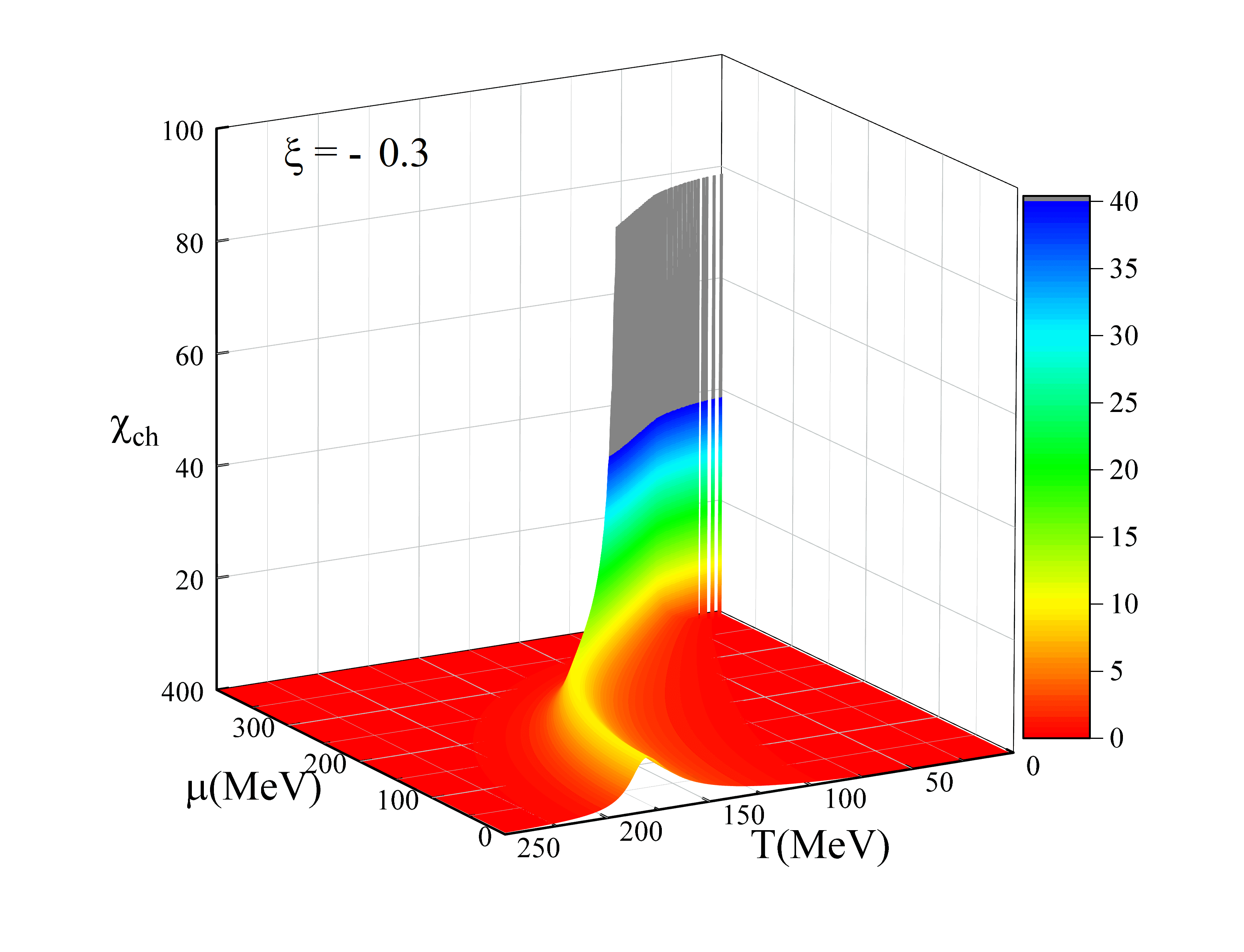}
 	\caption{\label{Fig:Chi_uT}  The 3-dimensional plot  of chiral susceptibility  $\chi_{ch}$  for $\xi=-0.3$  in the entire    $\mu$ and  $T$  ranges of interest. The gray  area means  $\chi_{ch}$ is divergent. (The values remain finite due to numerical problems (differential quotient). The peak height at high $\mu$ is two orders of magnitude higher than in small $\mu$ cases and can be considered "divergent".) }
 \end{figure}

 \begin{figure}
 \includegraphics[width=0.5\textwidth]{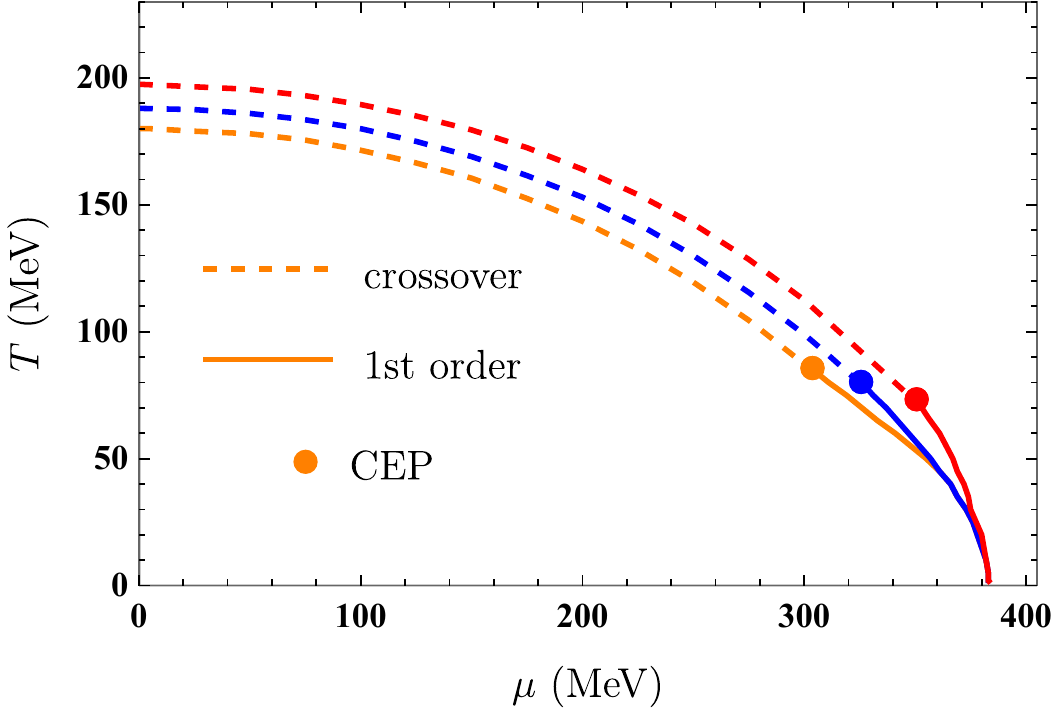}
 \caption{\label{Fig:CEP} 
 	 The chiral phase diagram for different anisotropy parameters in the NJL model. The solid lines denote the first-order phase transition curves, the dashed lines denote the crossover transition curves, and the solid dots represent the  CEPs. We observe that the CEP is shifted towards larger values of the quark chemical potential but smaller values of the temperature for higher anisotropy parameters.}
\end{figure}

\begin{figure}
\includegraphics[width=0.45\textwidth]{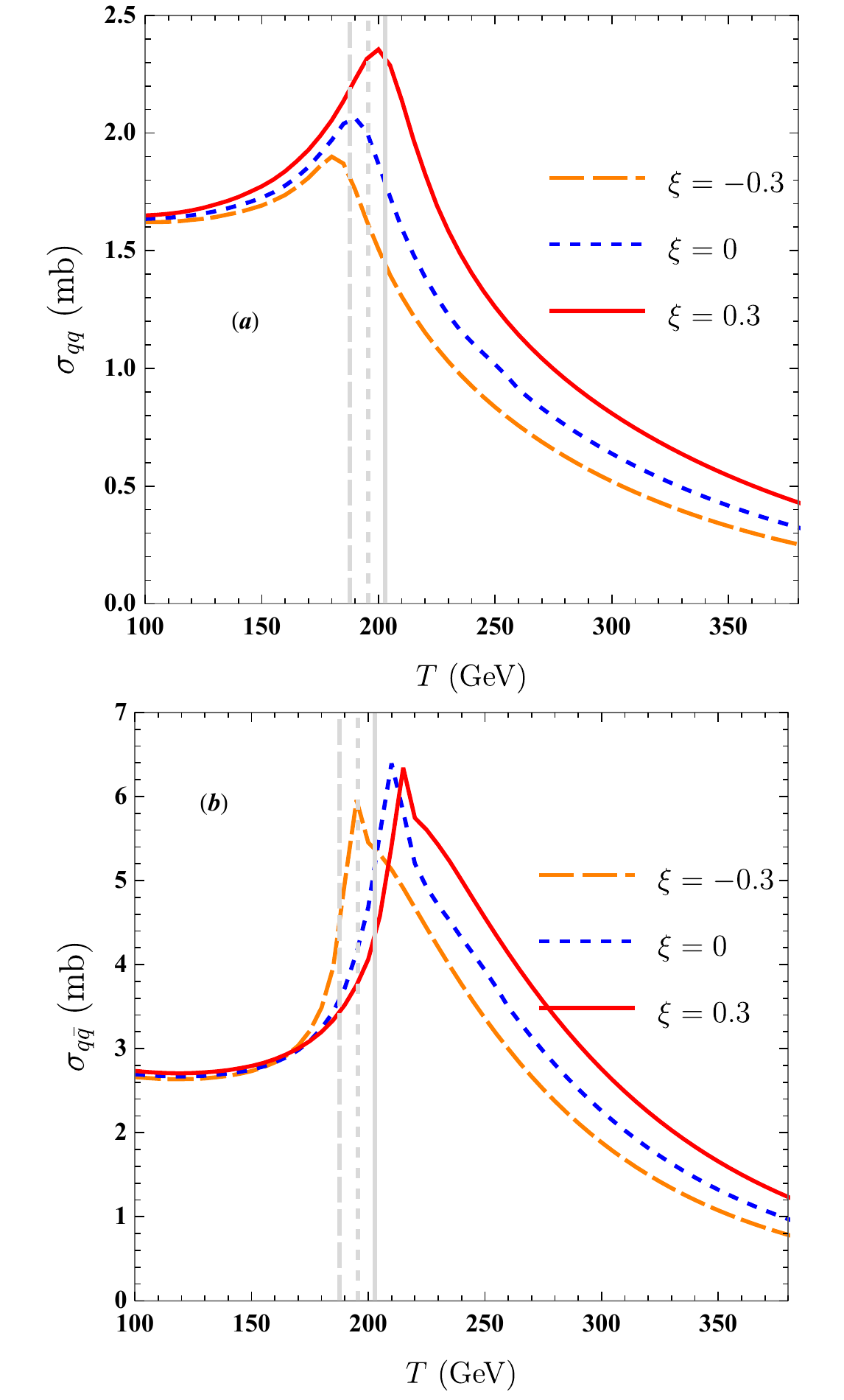}
\caption{\label{Fig:section} (plot $a$) The  cross-section of total   quark-quark scattering processes  $\bar{\sigma}_{qq}$ as a function of temperature at $\mu=0$~MeV for different anisotropy parameters.  (plot $b$)  The  cross-section of total quark-antiquark scattering processes  $\bar{\sigma}_{q\bar{q}}$   as a function of temperature at $\mu=0$~MeV for different anisotropy parameters, i.e., $\xi= -0.3$  (orange  broad dashed line), $\xi=0$ (blue dashed line),~ $\xi=0.3$  (red solid line).  The gray  vertical lines (from left to right) represent the critical temperatures $T_c=$ 180~MeV,~188~MeV,~197~MeV for $\xi=-0.3,~0,~0.3$.}
\end{figure}

\begin{figure}
\includegraphics[width=0.42\textwidth]{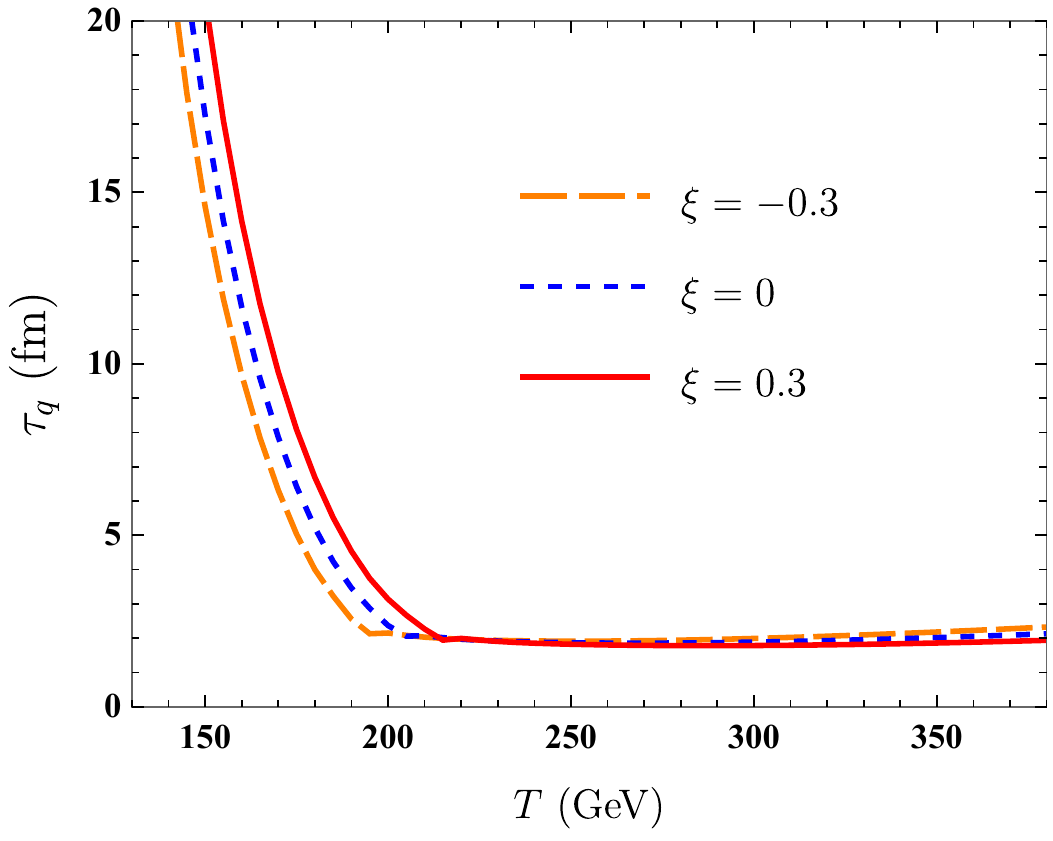}
\caption{\label{Fig:tau}The relaxation time of quark  at   $\mu=0$~MeV as a function of temperature for different anisotropy parameters, i.e., $\xi= -0.3$  (orange broad dashed line), $\xi=0$ (blue dashed line),~ $\xi=0.3$  (red solid line). } 
\end{figure}

 In this work, the chiral critical temperature, $T_{c}$, is determined  by  the peak location  of  the associated chiral susceptibility $\chi_{ch}$, which is defined as $\chi_{ch}=\big|\frac{dm_q}{dT}\big|$. 
 We stress that   the criterion of obtaining the  chiral critical  temperature is different in different papers, and there are some  shortcomings in the NJL model, such as   parameter ambiguity, nonrenormalization and the absence of gluonic dynamics, so the  value of $T_{c}$ in present work  is not expected to  quantitatively describe  the lattice QCD result. Anyway,  these cannot affect our present  qualitative results.
  The temperature dependence of chiral susceptibility $\chi_{ch}$  for different $\xi$  at $\mu=0$~MeV is plotted in  Fig.~\ref{Fig:Mass} (b). We observe that  $T_{c}$ exhibits a significant $\xi$ dependence. As $\xi$ increases, $T_{c}$ shifts toward higher temperatures and the height of  peak    decreases. The locations of $T_{c}$ for $\xi=-0.3,~0,~0.3$  are $\sim$ 180 MeV,~188 MeV,~197~MeV, which means a change of about 10$\%$ in temperature.    
  Actually,  the  in-medium meson masses also can be regarded as  a signature of  chiral phase transition. 
   In Fig.~\ref{Fig:Mass}(a),  we display  the variation of   $\pi$  and $\sigma$  meson masses  with   temperature for different $\xi$ at $\mu=0$~MeV. As can be seen that  the  $\pi$ mass  remains  approximately constant up to a particular temperature  whereas the   $\sigma$ mass first decreases  and then increases. As temperature increases further,  difference between $\pi$ mass and $\sigma$ mass also decreases and finally vanishes, at this time,  $\sigma$ and $\pi$ mesons are degenerate and  become unphysicals degrees of freedom, which indicates the  restoration of chiral symmetry. 
  And  before $\pi$ and $\sigma$ meson masses emerge, $\pi$  mass decreases as $\xi$ increase beyond $T_{c}$, whereas
 $\sigma$ mass first increases and then decreases with the increase of $\xi$. This qualitative behavior of meson masses with $\xi$ also is observed in our previous report~\cite{Zhang:2020zrv}  based on the  quark-meson model. Our result also slightly differs from  the finite size study of the NJL model~\cite{Deb:2020qmx},
which shows that below  the critical temperature,  $\pi$ mass enhances as   the system size decreases, while  $\sigma$   mass first remains unchanged then increases.
Furthermore, we see that  at a certain temperature, two times constituent quark mass ($2m_q$) is equal to   $\pi$  mass, the  pion meson is no longer a bound state but only a $q\bar{q}$ resonance and  obtains a finite decay width. Accordingly, the Mott transition temperature by the definition $m_{\pi}(T_{Mott})=2m_q(T_{Mott})$ can be obtained.  The Mott temperatures for $\xi=-0.3,~0$ and 0.3 turn out to be $\sim$ 187~MeV,~196~MeV and 206~$\mathrm{MeV}$, respectively, which is slightly higher than the corresponding $T_c$. In the vicinity of the Mott temperature, $\sigma$ meson features its minimal mass. In
Fig.~\ref{Fig:Meson} (b), we  illustrate  the variation of the decay widths of both  $\sigma$ and $\pi$ mesons with temperature for different $\xi$. As can be seen, 
 the decay width of $\sigma$ meson, $\Gamma_{\sigma}$, exists in the entire temperature range whereas  the decay width of $\pi$ meson, $\Gamma_{\pi}$, starts after the   Mott temperature.
  At high temperature,  the  merging behaviors of  decay widths for different mesons  are  also observed.
And with the increase of $\xi$,  the decay widths of mesons have a reduction.

We continue  the analysis in  finite quark chemical potential case to    investigate  the effect of momentum anisotropy on the  phase boundary  and the  CEP position.  First, we display the temperature- and quark chemical potential-dependence of constituent quark mass $m_q$ for different anisotropy parameters, as shown in Fig.~\ref{Fig:mq_uT}. We can see that  at small $\mu$, $m_q$ continuously decreases with increasing $T$, whereas
 $m_q$  has a  significant discontinuity or a sharp drop  along $T$-axis  at sufficiently high $\mu$,  which  is usually considered as the appearance of the  first-order phase transition.
 To visualize the phase diagram we use the significant divergency of $\chi_{ch}$  at sufficiently high chemical potential as the criterion for a first-order phase transition, as shown in Fig.~\ref{Fig:Chi_uT}.  With the decrease of $\mu$, the first-order phase  transition terminates at a critical endpoint (CEP), where the phase transition is expected to be  of second order. As  $\mu_{}$ decreases further,  the  maximum of the  chiral susceptibility ($\chi_{ch}$)  as the crossover criterion. 
    The full    chiral boundary lines in the ($\mu$-$T$) plane for three different values of  $\xi$ are displayed in Fig.~\ref{Fig:CEP}. 
 We  observe that   as the increase of $\xi$, the phase boundary  shifts    toward higher quark chemical potentials and higher temperatures.    We  also can see the CEP position  is sensitive to the variation of $\xi$. 
  As $\xi $ increases, the location of CEP shifts to higher $\mu$ and smaller $T$.
    The CEP location ($T_{CEP},~\mu_{CEP}$) in this work are separately  presented  at  (298.70~$\mathrm{MeV}$, 88.2~$\mathrm{MeV}$),~(321.8~$\mathrm{MeV}$,~82.4~$\mathrm{MeV}$),~(348.4~$\mathrm{MeV}$,~74.2~$\mathrm{MeV}$) for $\xi=-0.3,~0,~0.3$. The the position of  CEP for $ \xi=0$  in this work is almost consistent with the  existing result~\cite{Chaudhuri:2019lbw}  in the same parameter fit.
    The value of $\mu_{CEP}$ ($T_{CEP}$) from $\xi=-0.3$ to $\xi=0.3$   increases (decreases)  by approximately 16$\%$ (17$\%$), in other words,  the influence degree  of momentum anisotropy on temperature of  CEP is almost the same as that on the quark chemical potential of CEP. This is different to  the result of  Ref.~\cite{Zhang:2020zrv}, which has shown that in the  quark meson model the impact  of momentum anisotropy on the quark chemical potential of the CEP is significantly dominant than that on the  associated temperature.
  In addition, in the study of finite volume effect, the result of Ref.~\cite{Bhattacharyya:2012rp} has indicated that in the  PNJL model the finite volume affects the CEP shift along the temperature more strongly  than along the quark  chemical  potential shift.  
 And when the system size is reduced to 2 fm, the CEP in the PNJL model vanishes and the whole chiral phase boundary becomes a crossover curve.
  Based on this result, there also exists a possibility that if  $\xi$ further increases, the CEP may disappear from the phase diagram. 
  
\begin{figure}
	\includegraphics[width=0.45\textwidth]{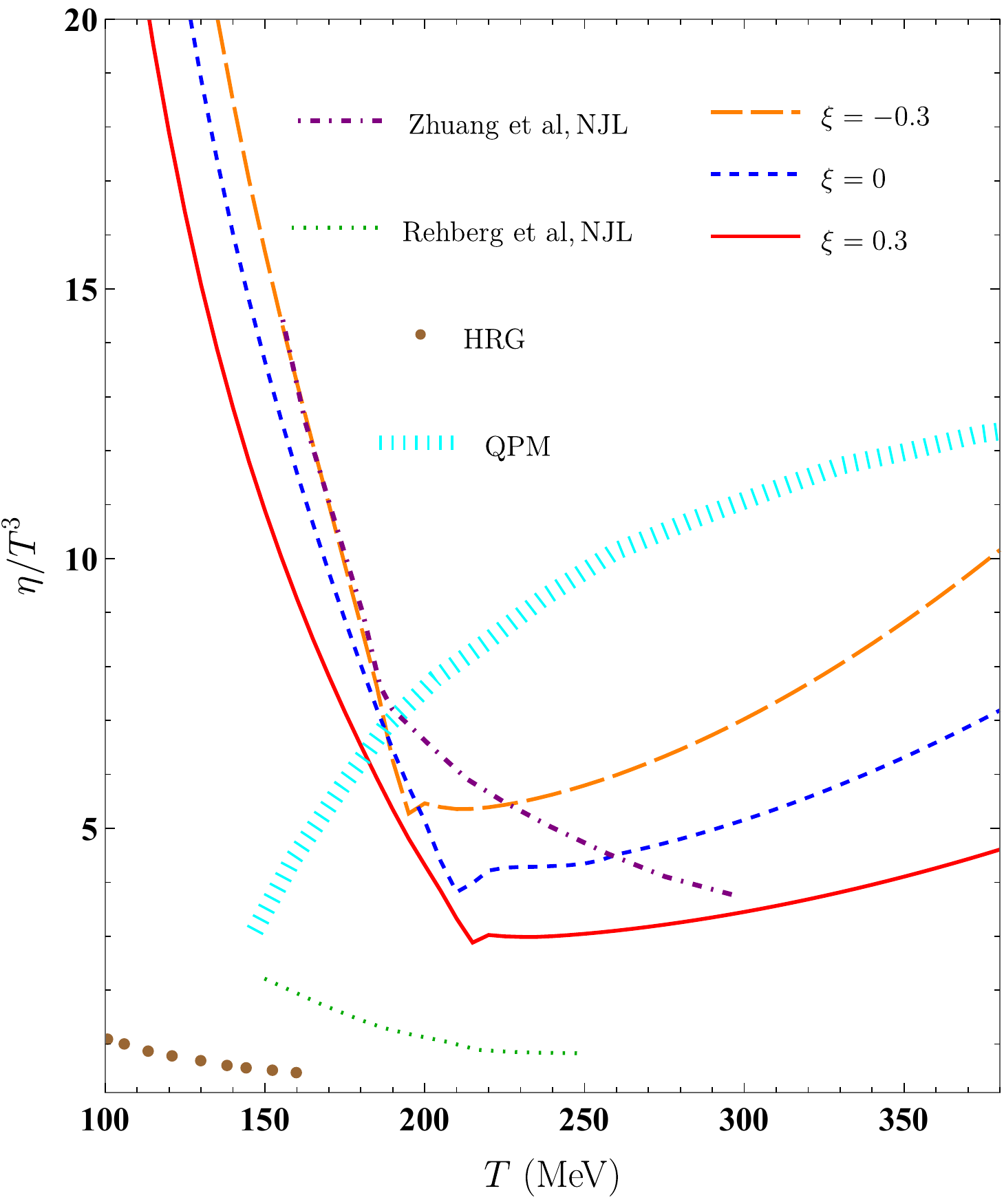}
	\caption{\label{Fig:viscosity}The temperature dependence of    scaled shear viscosity $\eta/T^3$ in quark matter at  vanishing chemical potential for different anisotropy parameters i.e., $\xi=$ $-0.3$ (orange  broad dashed line), $\xi=0$ (blue dashed line),~and $\xi=0.3$ (red solid line). The thick cyan dotted line represents the result in the $N_{f}=3$ quasiparticle model (QPM)~\cite{shear-quasi2}, which is an effective model for the description of non-perturbative QCD. The purple dotdashed line shows the result obtained in the   $N_{f}=2$ NJL model by Zhuang $et~al$~\cite{Zhuang:1995uf}. The brown dots show the result from hadron resonance gas (HRG) model~\cite{shear and electrical}. The green  dots correspond  to  the result of  Rehberg $et~al$ in the  $N_{f}=3$ NJL model~\cite{Rehberg:1996vd} using the averaged transition rate  method for the estimation of relaxation time.}
\end{figure}

\begin{figure}
	\includegraphics[width=0.45\textwidth]{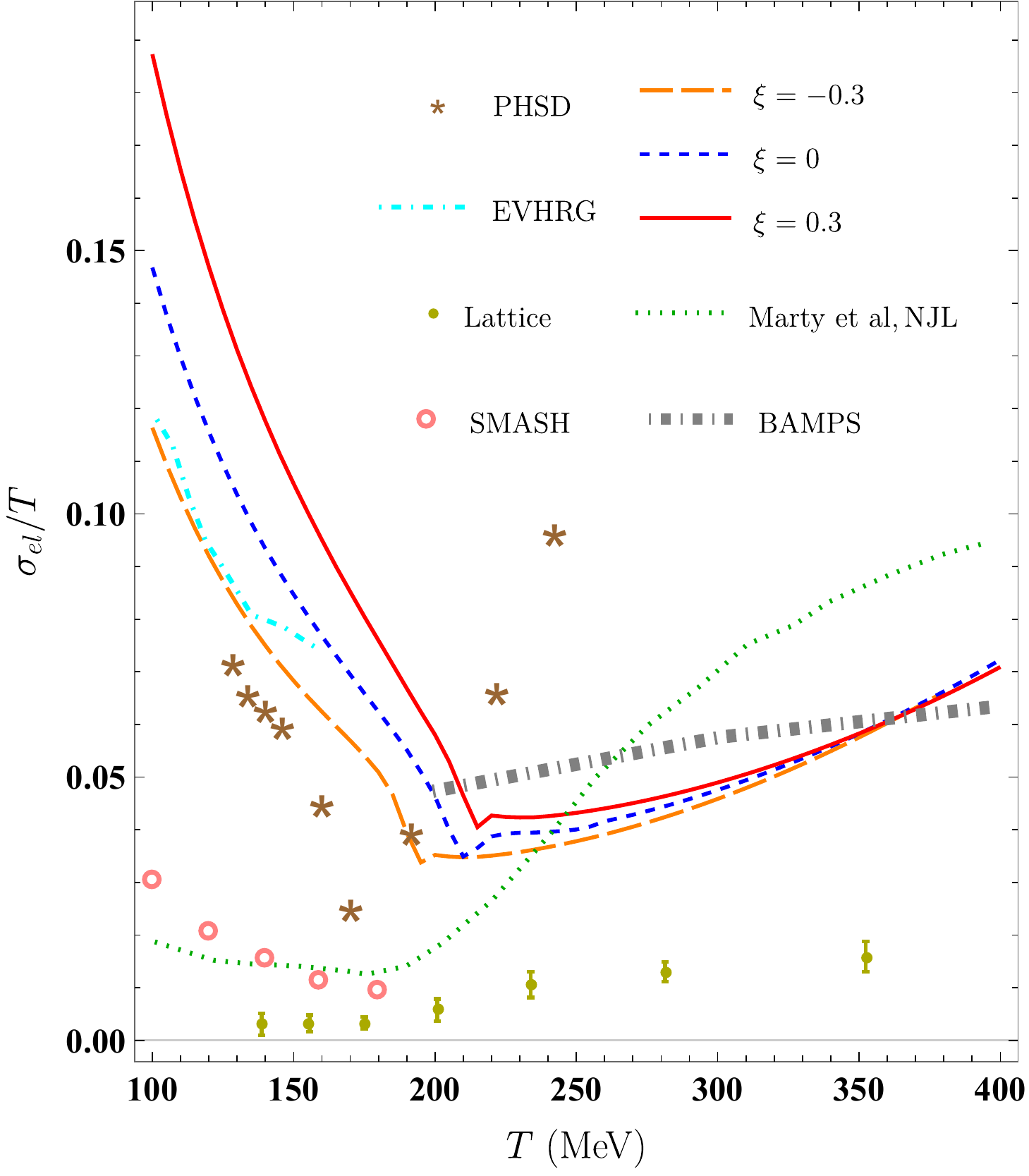}
	\caption{\label{Fig:conductivity}The temperature dependence of   scaled electrical conductivity $\sigma_{el}/T$  in quark matter at  vanishing chemical potential  for different anisotropy parameters i.e., $\xi=$ $-0.3$ (orange broad dashed line), $\xi=0$ (blue dashed line),~and $\xi=0.3$ (red solid line). The green dotted line shows  the result of  Marty $et ~al$ in $N_{f}=3$ NJL model~\cite{Marty:2013ita}. The thick gray dotdashed line represents the result  from the pQCD-based microscopic   Boltzmann Approach to Multi-Parton Scatterings  (BAMPS)  transport model~\cite{Greif:2014oia} with running coupling constant. 
		 The brown stars present the result in the Parton-Hadron-String Dynamics (PHSD) transport approach~\cite{Steinert:2013fza}.
		 The cyan dotdashed line shows the result within excluded volume hadron resonance gas (EVHRG) model with the RTA~\cite{electrical-EVHRG}. The darkyellow dots are the lattice date obtained from Ref.~\cite{Amato:2013naa}.  The red open circles are the calculation  for hadronic gas in the  transport approach- Simulating Many Accelerated Strongly-interacting Hadrons (SMASH)~\cite{Hammelmann:2018ath} based on  Green-Kubo formalism.}
\end{figure}

\begin{figure}
	\includegraphics[width=0.45\textwidth]{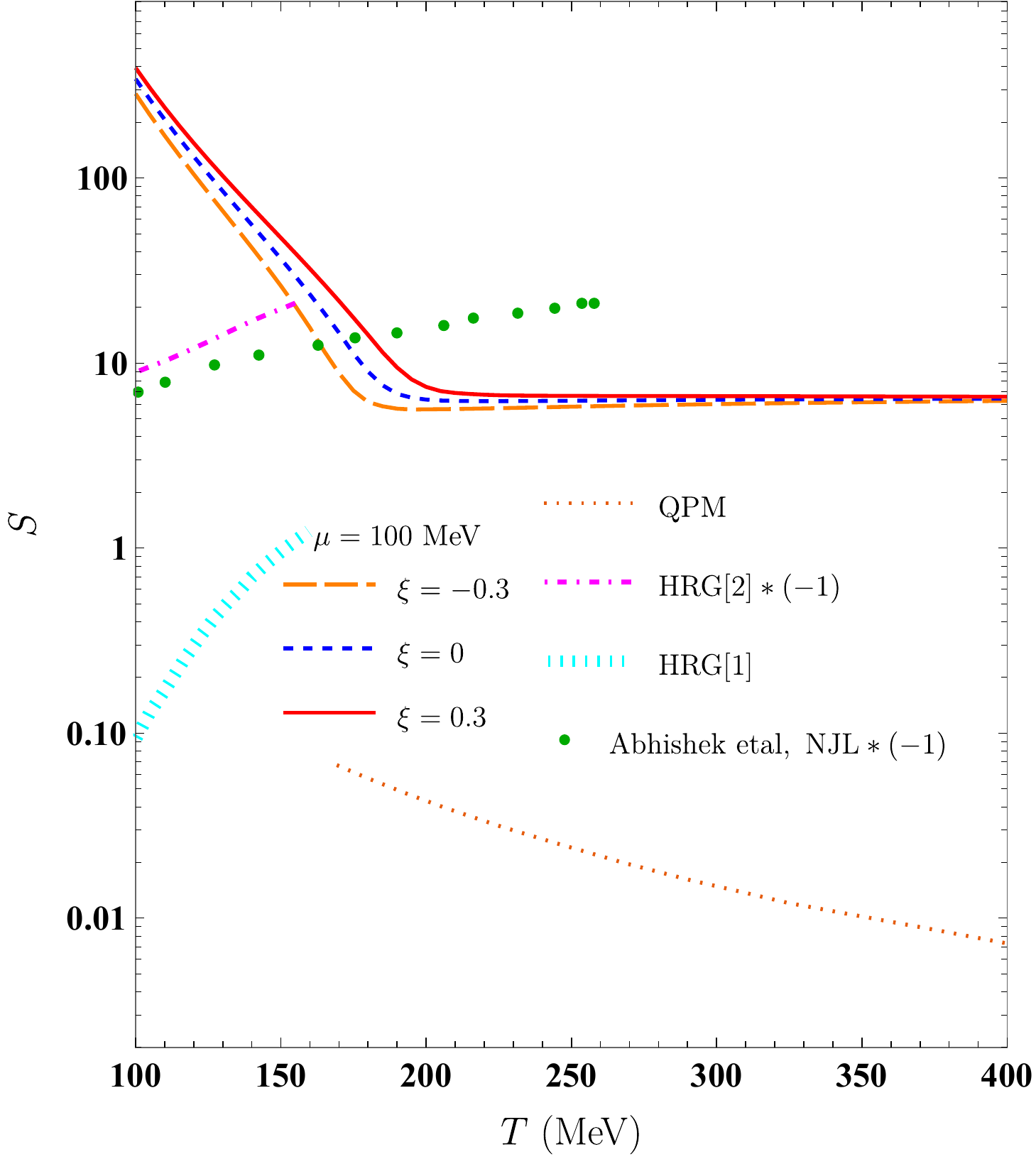}
	\caption{\label{Fig:Seebeck}The temperature dependence of   Seebeck coefficient in quark matter  at  $\mu=100$~MeV  for different anisotropy parameters i.e., $\xi=$ $-0.3$ (orange  broad dashed line), $\xi=0$ (blue dashed line),~and $\xi=0.3$ (red solid line). The brown dotted line corresponds to  the result for  the QGP in the quasiparticle model~\cite{Dey:2020sbm} at  $\mu_{q}=50$~MeV. The  cyan  thick-dotted line represents the result in hadron resonance gas model for $\mu_{B}=0.1$~GeV~\cite{Bhatt:2018ncr},  The mauve dotdashed line and green dots, respectively, represent  the results in HRG model for $\mu_{B}=50$~MeV~\cite{Das:2020beh} and the $N_{f}=2$ NJL model for $\mu=100$~MeV~\cite{Abhishek:2020wjm}, where  the gradient of  quark chemical potential apart from a spatial gradient in temperature  also is included. }
\end{figure}
To better understand  the qualitative behavior of   transport coefficients, we first discuss the results of the scattering  cross-sections and  the relaxation time.
In Fig~\ref{Fig:section},  we display  the   cross-section of  total  quark-quark    scattering processes   $\bar{\sigma}_{qq}=\bar{\sigma}_{uu\rightarrow uu}+\bar{\sigma}_{ud\rightarrow ud}$ (plot $a$) and  the  cross-section of total quark-antiquark   processes    $\bar{\sigma}_{q\bar{q}}=\bar{\sigma}_{u\bar{u}\rightarrow u\bar{u}}+\bar{\sigma}_{u\bar{d}\rightarrow u\bar{d}}+\bar{\sigma}_{u\bar{u}\rightarrow d\bar{d}}$  (plot $b$) as  functions of temperature at  different anisotropy parameters for vanishing quark chemical potential. 
As can be seen, $\bar{\sigma}_{qq}$  and $\bar{\sigma}_{q\bar{q}}$    have similar peak features in entire temperature region of interest. 
 More exact, the scattering cross-sections first increase, reaches a peak,  and decreases with increasing temperature afterwards. And  the  magnitude  of  $\bar{\sigma}_{q\bar{q}}$  is higher than that of  $\bar{\sigma}_{qq}$. This   is mainly due to  that the $s$ channel allows for a resonance of the exchanged meson with the incoming quarks, which leads to a large peak in the cross-section~\cite{Soloveva:2020hpr}. 
 We can also see that the scattering cross-sections in the weakly  anisotropic medium keep the same behaviors as those in the isotropic medium.
 As $\xi$ increases, $\bar{\sigma}_{qq}$ increases in the  entire temperature  domain of considered, whereas $\bar{\sigma}_{q\bar{q}}$  first decreases as $\xi$ increases and then increases as $\xi$ increases. 
  With an increase in $\xi$, the maximum of the scattering  cross-section shifts toward higher temperatures. The location of maximum for $\bar{\sigma}_{qq}$ at different $\xi$ is nearly in agreement with respective $T_{c}$. While the  peak positions of  $\bar{\sigma}_{q\bar{q}}$  respectively locate at $\sim 1.07~T^{-0.3}_{c}$, $1.07~T^0_{c}$, $1.10~T^{0.3}_{c}$ for $\xi=-0.3,~0,~0.3$ with $T^{\xi}_{c}$ denoting the chiral critical temperature for a fixed $\xi$.
  
The   dependence  of total  quark relaxation time $\tau_{q}$ on  temperature for vanishing quark chemical potential at different $\xi$ is displayed in Fig.~\ref{Fig:tau}. As can be seen,   $\tau_{q}$ first  decreases sharply with increasing temperature,  after an inflection point ($viz$, the peak position of $\bar{\sigma}_{q\bar{q}}$), $\tau_q$ is  modestly changing with temperature.  And the  increase  of     $\tau_{q}$  with $\xi$  is significant at low temperature  whereas at  high  temperature the reduction of $\tau_{q}$  with $\xi$ is imperceptible. This is the result of the competition between the quark number density and the total scattering cross section in Eq.~(\ref{eq:taulight}). At small temperature, the $\xi$ dependence of $\tau_{q}$ is mainly determined by  the inverse  quark number density  whereas at high temperature it  is primarily governed by the  inverse total cross-section  even though this effect is largely cancelled out by the inverse quark density effect.

Next, we are going to discuss the results   regarding various transport coefficients.
In Fig.~\ref{Fig:viscosity},  the temperature dependence of  scaled  shear viscosity $\eta/T^3$ in quark matter for different momentum anisotropy parameters at a vanishing chemical potential is displayed. 
We observe that with increasing temperature, $\eta/T^3$ first decreases, reaches a  minimum around the critical temperature, and  increases afterwards. The temperature  position of  minimum for $\eta/T^3$  is consistent  with  the  temperature of  peak for $\bar{\sigma}_{q\bar{q}}$. This  dip structure of $\eta/T^3$  can mainly  depend on   the result of a competition between  quark distribution function $f^0_{q}$  and  quark relaxation time $\tau_{q}$ in the  integrand of  Eq.~(\ref{eq:shear}). 
 The increasing feature of $\eta/T^3$ in  low temperature domain is governed by $\tau_{q}$, while in high temperature domain the increasing  behavior of $f^0_{q}$ overwhelms the decreasing behavior of $\tau_{q}$, leading     $\eta/T^3$  become an increasing function of  temperature. 
  Furthermore, we observe that  as an increase in
  $\xi$, $\eta/T^3$  has an overall  enhancement and  the  minimum  of  $\eta/T^3$ shifts to higher temperatures. 
  The location of the  minimum for  $\eta/T^3$   at different $\xi$  is   consistent with the peak position of $\bar{\sigma}_{q\bar{q}}$. 
  And we observe that $\eta/T^3$ decreases as $\xi $ increases in entire temperature region. 
 we also compare our result for $\xi=0$ with  the results reported in other previous literature.  
  The calculation of $\eta/T^3$ in hadron resonance gas (HRG) model~\cite{shear and electrical} (brown dots) using the RTA is  a decreasing function with temperature, which  is qualitatively similar to ours below the  critical temperature.
  The quantitative difference between HRG model result and ours  can be attributed to the uses of various degrees of freedom and the difference of scattering cross-sections.
  The result of Zhuang $et~ al$~\cite{Zhuang:1995uf} in the  $N_{f}=2$ NJL model (purple dotdashed line)  is  of  the same  order of   magnitude  as ours, while  at high temperature their result still remains a decreasing feature because an ultraviolet cutoff is used in all momentum integral  whether  temperature is finite or zero.    The result estimated in the quasiparticle (QPM)~\cite{shear-quasi2}  is  a logarithmically  increasing function of temperature  beyond the critical temperature, and is quantitively larger than ours beyond critical temperature due to the differences in both the effective mass of quark and the relaxation time.   The result of Rehberg $et~al$~\cite{Rehberg:1996vd} for  the  $N_{f}=3$ NJL model in the temperature regime  close to the critical temperature is  smaller than ours, and the obvious  dip structure is not  observed  because  the momentum cutoff is also used at finite temperature.

In Fig.~\ref{Fig:conductivity},  we plot the thermal  behavior of  scaled electrical conductivity $\sigma_{el}/T$  at $\mu=0$~MeV for different $\xi$. Similar to  the temperature dependence of $\eta/T^3$,   $\sigma_{el}/T$  also exhibits a dip structure in the  entire temperature region of interest. 
 We also present the comparison with other previous results.
 The result obtained from the  PHSD approach~\cite{Steinert:2013fza} (brown stars), 
 where the plasma evolution is solved by a Kadanoff-Baym type equation,
  also has a valley structure, eventhough the  location of the   minimum is different with ours.
 We also observe that in the temperature region dominated by hadronic phase,  the thermal behavior  of $\sigma_{el}/T$ using the microscopic simulation code  SMASH~\cite{Hammelmann:2018ath} (pink open circles)  is  similar to ours. 
 Furthermore, our result is  much larger than the lattice QCD data (darkyellow dots) taken from Ref.~\cite{Amato:2013naa} due to the uncertainity in the parameter set and the uninclusion of gluonic dynamics.
    The result within exclude volume hadron resonance gas (EVHRG) model\cite{electrical-EVHRG} (cyan dotdashed line) and the result  obtained from partonic cascade BAMPS~\cite{Greif:2014oia} (gray thick dotdashed line) are in the both  qualitative and quantitive   similar to our calculations below the critical temperature  and  beyond the critical temperature, respectively.
    Our shape is similar to the result of  Marty $et~al$  obtained within the  $N_{f}=3$ NJL model~\cite{Marty:2013ita} (green dotted line),  the numerical discrepancy mainly comes from  the differences in both the  parameter set  and the normalization of the scattering cross-section.
  In addition, $\sigma_{el}/T$  shows a  different $\xi$ dependence than $\eta/T^3$.  More exact, $\sigma_{el}/T$  first increases as $\xi$  whereas, as $T$ increases further,  the values $\sigma_{el}/T$  for  different $\xi$ gradually approach and eventually overlap, which is  different to the result in Ref.~\cite{Srivastava:2015via}.   In Ref.~\cite{Srivastava:2015via}, $\sigma_{el}/T$ of the QGP is a monotonic  increasing function of  $\xi$ because  the effect of momentum anisotropy is not incorporated in the calculation of the relaxation time and the effective mass of  quasiparticles,  the  $\xi $ dependence of $\sigma_{el}/T$  is only determined  by the  anisotropic distribution function.
  We also observe that with the increase of $\xi$, the minimum of $\sigma_{el}/T$ shifts to higher temperatures, which is similar to $\eta/T^3$, however, the height of the minimum increase, which is opposite to $\eta/T^3$.
   
Finally, we study  Seebeck coefficient $S$ in quark-antiquark matter.  Due to  the sensitivity of $S$ to charge type of particle species,  at a vanishing chemical potential,  quark number density $n_q$ is equal to  antiquark  number density $n_{\bar{q}}$,  the contribution of quarks to $S$ is exactly compensated with the counterpart of antiquarks.  Thus,  a finite quark chemical potential is required to obtain a non-zero thermoelectric current in the medium.
In Fig.~\ref{Fig:Seebeck}, we plot the variation  of $S$ with respect to temperature
for different $\xi$ at $\mu=100$~MeV. The comparison with other previous calculations, which are all   performed in the kinetic theory under the RTA,  also is presented.  We remind the reader that at a finite $\mu$, $n_q$ is larger than  $n_{\bar{q}}$, the contribution of quarks to  total $S$ in magnitude is always prominent.
 As shown in Fig.~\ref{Fig:Seebeck},  the sign of  $S$ in our investigation is positive, which indicates that the dominant carriers of converting heat gradient to the electric field is positively charged quarks, i.e., $up$ quarks. Actually,  the positive or negative of $S$ is  mainly  determined by the factor $(E_q-\mu_q)$ in the integrand of Eq.~(\ref{eq:Seebeck}). 
  In Ref.~\cite{Dey:2020sbm}, Seebeck coefficient  studied in the QPM (brown dotted line)  at  $\mu_{}=50$~MeV also exhibits  a decreasing feature with increasing temperature. 
 The  result of Abhishek $et~al$~\cite{Abhishek:2020wjm} at $\mu=100~$MeV in the $N_{f}=2$ NJL model (the green dots) is much  different with ours. 
 In Ref.~\cite{Abhishek:2020wjm}, $S$ is negative and its absolute value exhibits  an increasing function with temperature. The reasons behind this quantitative and qualitative  discrepancy are twofold: (1) the relaxation  time in Ref.~\cite{Abhishek:2020wjm} is estimated by using the averaged transition rate $\bar{w}_{ij}$ while   our  relaxation time   is obtained from the thermally averaged  cross-section of elastic scattering (the detailed comparison of  two methods can be found in Ref.~\cite{Soloveva:2020hpr});
 (2) in Ref.~\cite{Abhishek:2020wjm}, the spatial  gradient of   chemical potential also is included apart from the  temperature gradient, accordingly the sign of $S$ is mainly determined by  a  factor $(E_q-\omega/n_q)$ with $\omega$ denoting the enthalpy density in the  associated formalism. Due to the single-particle energy $E_q$ remains smaller than $(\omega/n_q)$,  $S$ in Ref.~\cite{Abhishek:2020wjm} is negative. We also see that  with increasing temperature, $S$  sharply decreases below $T_{c}$, whereas  the decreasing feature of $S$ is unconspicuous above $T_{c}$.  And the value of $S$ at low $T$  is much larger than that at high $T$. This also is different to the result in   Ref.~\cite{Abhishek:2020wjm}, where  the  absolute  value of $S$  in quark matter increases with increasing temperature  because of  the increasing behaviors of both   the factor $|-\omega/n_{q}|$ and the equilibrium distribution function. 
  In addition, Seebeck coefficient in HRG model~\cite{Das:2020beh,Bhatt:2018ncr} is also positive (negative) without (with) the spatial gradient of chemical potential (cyan  thick dotted line and  mauve  dotdashed line).  Nevertheless, the absolute value of $S$ in hadronic matter is still an increasing function of  temperature regardless of  the spatial gradient of $\mu$.  
 We also see that   as $\xi$ increases, $S$ has a quantitative enhancement, which is mainly due to a significant  rise in the  thermoelectric conductivity $\alpha$, eventhough  $1/\sigma_{el}$ has a cancellation effect on the increase of $S$. At sufficiently  high temperature,  the rise in  $1/\sigma_{el}$ can  almost compensate  with the reduce of $\alpha$, as a result,   $S$ varies unsignificantly
  with  $\xi$ of interest, compared to the value of $S$ itself. 

\section{Summary}\label{sec:sum}
We  phenomenologically  investigated  
the impact of weak  momentum-space anisotropy on the chiral phase structure, mesonic properties, and  transport properties in  the 2-flavor NJL model.  The momentum anisotropy, which is  induced by initial preferential expansion of  created   fireball in heavy-ion collisions  along the beam direction,    can be incorporated in the calculation through the parameterization of  anisotropic  distribution function.
Our result has shown that  the chiral phase  transition  is a smooth  crossover for vanishing  quark chemical potential, independent of  anisotropy parameter $\xi$, and an   increase in $\xi$ even  can hinder the restoration of the chiral symmetry.
We found the CEP  highly sensitive to the  change in $\xi$.  With the increase of $\xi$,  the CEP shifts to higher $\mu$ and smaller $T$, and the momentum anisotropy affects the CEP temperature to almost the same degree as it affects the CEP chemical potential.  Before  the merge of $\pi$ and $\sigma$ meson masses, the $\xi$ dependence of $\pi$ meson mass   is opposite to that  of $\sigma$ meson mass.

 We also  studied the thermal behavior of   various transport coefficient, such as scaled shear viscosity $\eta/T^3$, scaled electrical conductivity $\sigma_{el}/T$ and  Seebeck coefficient $S$ at  different $\xi$. The associated $\xi$-dependent expressions are derived by solving  the relativistic Boltzmann-Vlasov transport equation in the relaxation time approximation, and the momentum anisotropy effect also is embedded in the estimate of  relaxation time.
  We found $\eta/T^3$ and $\sigma_{el/T}$  have a dip structure around the critical temperature.
  Within the consideration of momentum anisotropy, $\eta/T^3$  decreases as $\xi$ increases  and the minimum shifts to higher temperatures.
  As the increase of $\xi$, $\sigma_{el}/T$ significantly increases for low temperature whereas  the sensitivity of  $\sigma_{el}/T$ to $\xi$  for high temperature is greatly reduced, which is different from the behavior of  $\eta/T^3$  with $\xi$.
  We also  found the sign of $S$ at $\mu=100$~MeV in present work is positive, indicating the dominant carriers  for converting the  thermal gradient to the electric field are  $up$ quarks. 
 And  with increasing temperature, $S$  first decreases sharply then almost flattens out.  At low temperature, $S$ significantly increases with an increase of $\xi$, whereas at high temperature the rise  is marginal compared to the value of  $S$ itself.

We note it is of strong interest to include the Polyakov-loop potential in the present  model  to study  both  chiral and confining dynamics in an anisotropic quark matter. And a more realistic   ellipsoidal momentum anisotropy  characterized by two independent anisotropy parameters can be applied to gain a deeper understanding of the QGP properties.  In present work,  there is no any proper time dependence has been given to the anisotropy parameter. However, in the  realistic case, $\xi$ varies with the proper time starting from the initial proper time up to a time when the system becomes isotropic. Thus, we also can introduce  a proper time dependence to the anisotropy parameter\cite{Martinez:2008di} to better explore the effect of time-dependent momentum anisotropy on chiral phase transition.
In addition,  the investigation of  the thermoelectric coefficients  specially the magneto-Seebeck coefficient and Nernst coefficient in magnetized quark matter based the  PNJL model would be an attractive direction, and we may work on it in the near future. 

\acknowledgments
This research is supported by Guangdong Major Project of Basic and Applied Basic Research No. 2020B0301030008, Natural Science Foundation of China with Project No. 11935007. The authors thank the anonymous referee for the constructive inputs and suggestions.
\appendix

\section*{Appendix}
In the $N_{f}=2$ NJL model, there are 12 different elastic scattering processes:
\begin{eqnarray}
u\bar{u}\rightarrow u\bar{u},~u\bar{d}\rightarrow u\bar{d},~u\bar{u}\rightarrow d\bar{d},\nonumber\\
uu\rightarrow uu,~ud\rightarrow ud,~\bar{u}\bar{u}\rightarrow \bar{u}\bar{u},\nonumber\\
\bar{u}\bar{d}\rightarrow \bar{u}\bar{d},~d\bar{d}\rightarrow d\bar{d},~d\bar{d}\rightarrow u\bar{u},\nonumber\\
d\bar{u}\rightarrow d\bar{u},~dd\rightarrow dd,~\bar{d}\bar{d}\rightarrow \bar{d}\bar{d}.
\end{eqnarray}
The  explicit expressions of the matrix elements  squared for  $u\bar{u}\rightarrow u\bar{u}$, $u\bar{d}\rightarrow u\bar{d}$ and $ud\to ud$ processes via  exchange of scalar and/or pseudoscalar mesons to $1/N_{c}$ order are given as 
\begin{widetext}
\begin{eqnarray}
|\bar M_{u\bar u\rightarrow u\bar u}|^2(s,t) &=&
s^2|D^\pi_s|^2
+t^2|D^\pi_t|^2+
(s-4m^2)^2|D^\sigma_s|^2
+(t-4m^2)^2|D^\sigma_t|^2+\frac{1}{N_c}Re\bigg[st D^{\pi^*}_sD^\pi_t
+s(4m^2-t)D^{\pi^*}_sD^\sigma_t\nonumber\\
&&+t(4m^2-s)D^\pi_tD^{\sigma^*}_s
+(st+4m^2(s+t)-16m^4)D^\sigma_tD^{\sigma^*}_s\bigg],\\
|\bar M_{u\bar d\rightarrow u\bar d}|^2 (s,t) &=&
4s^2|D^\pi_s|^2
+t^2|D^\pi_t|^2
+(t-4m^2)^2|D^\sigma_t|^2-\frac{1}{N_c}Re\bigg[2st D^{\pi^*}_sD^\pi_t
+2s(4m^2-t)D^{\pi^*}_sD^\sigma_t\bigg],\\
|\bar M_{u d\rightarrow u d}|^2 (t,u) &=&
4u^2|D^\pi_u|^2
+t^2|D^\pi_t|^2
+(t-4m^2)^2|D^\sigma_t|^2-\frac{1}{N_c}Re\bigg[2ut D^{\pi^*}_sD^\pi_t
+2u(4m^2-t)D^{\pi^*}_uD^\sigma_t\bigg].
\end{eqnarray}
\end{widetext}
The meson propagators in the above are $\xi$-dependent.
Based on above formulae of three scattering processes, the  matrix element squared for remaining  scattering processes can be  obtained through charge conjugation and crossing symmetry~\cite{Friesen:2013bta,Zhuang:1995uf}.

\end{document}